\documentclass[pra,superscriptaddress,reprint]{revtex4-1}
\usepackage{graphicx}
\usepackage{bm}
\usepackage{ulem}
\usepackage{amsmath}
\usepackage{amssymb}
\usepackage{hhline}

\makeatletter
\renewcommand{\fnum@figure}{\textbf{Figure \thefigure}}
\makeatother

\begin{document}

\title{Hybrid quantum systems based on magnonics}

\author{Dany Lachance-Quirion}
\email{dany.lachance.quirion@qc.rcast.u-tokyo.ac.jp}
\author{Yutaka Tabuchi}
\author{Arnaud Gloppe}
\author{Koji Usami}
\affiliation{Research Center for Advanced Science and Technology (RCAST), The University of Tokyo, Meguro-ku, Tokyo 153-8904, Japan}
\author{Yasunobu Nakamura}
\email{yasunobu@ap.t.u-tokyo.ac.jp}
\affiliation{Research Center for Advanced Science and Technology (RCAST), The University of Tokyo, Meguro-ku, Tokyo 153-8904, Japan}
\affiliation{Center for Emergent Matter Science, RIKEN, Wako, Saitama 351-0198, Japan}

\date{February 19, 2019}

\begin{abstract}
Engineered quantum systems enabling novel capabilities for communication, computation, and sensing have blossomed in the last decade. Architectures benefiting from combining distinct and complementary physical quantum systems have emerged as promising platforms for developing quantum technologies. A new class of hybrid quantum systems based on collective spin excitations in ferromagnetic materials has led to the diverse set of experimental platforms which are outlined in this review article. The coherent interaction between microwave cavity modes and collective spin-wave modes is presented as the backbone of the development of more complex hybrid quantum systems. Indeed, quanta of excitation of the spin-wave modes, called magnons, can also interact coherently with optical photons, phonons, and superconducting qubits in the fields of cavity optomagnonics, cavity magnomechanics, and quantum magnonics, respectively. Notably, quantum magnonics provides a promising platform for performing quantum optics experiments in magnetically-ordered solid-state systems. Applications of hybrid quantum systems based on magnonics for quantum information processing and quantum sensing are also outlined briefly.
\end{abstract}

\maketitle

\section{Introduction}
\label{sec:introduction}

In the last few decades, technologies engineered from quantum systems have flourished and found applications in quantum computing~\cite{Feynman1986,Nielsen2001,Ladd2010}, quantum simulation~\cite{Feynman1982,Aspuru-Guzik2005,Cirac2012}, quantum communication~\cite{Duan2001,Kimble2008,Reiserer2014a,Wehner2018}, and quantum sensing~\cite{Degen2017}. An exciting strategy has been explored in the last few years to access through hybrid quantum systems novel capabilities in existing quantum technologies by the combination of distinct physical systems possessing complementary characteristics~\cite{Xiang2013,Kurizki2015,Schleier-smith2016}. For example, merging techniques developed in the field of circuit quantum electrodynamics (cQED) to the field of electron spin resonance has enabled drastic improvements of the detection sensitivity of electron spins~\cite{Bienfait2015,Bienfait2017,Eichler2017,Probst2017,Morton2018}.

\begin{figure}\begin{center}
\includegraphics[scale=1]{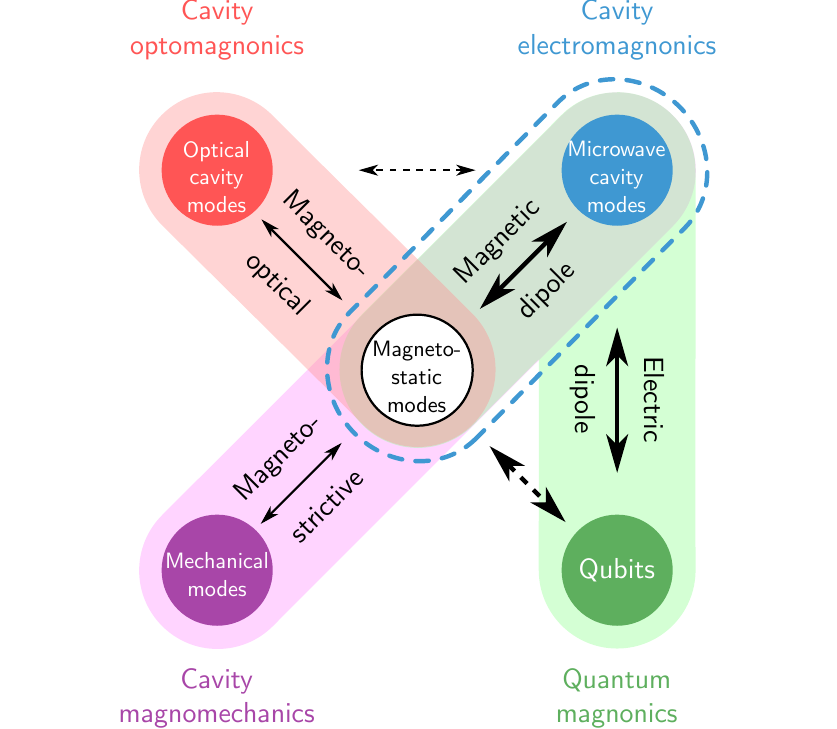}
\caption{\textbf{Hybrid quantum systems based on magnonics.}
Schematic diagram of the different interactions in hybrid quantum systems based on quanta of collective spin excitations of magnetostatic modes in a ferromagnetic crystal. The basis of most implementations of hybrid systems with magnons is cavity electromagnonics, where a strong coupling between magnetostatic modes and microwave cavity modes is achieved through a collective enhancement of the magnetic dipole interaction of a single spin. In quantum magnonics (green), the strong electric dipole coupling of superconducting qubits to microwave cavity modes is used to engineer an effective strong coupling between the magnetostatic modes and the qubits. In cavity optomagnonics (red), magneto-optical effects allow a coupling between the magnetostatic modes and optical cavity modes, enabling, in combination with cavity electromagnonics, bidirectional conversion between optical and microwave photons. In cavity magnomechanics (purple), the magnetostrictive force couples the magnetostatic modes and mechanical modes of the ferromagnetic crystal. Strong (weak) interactions are shown as bold (thin) arrows and direct (indirect) interactions are shown as solid (dashed) arrows.
\label{fig:hybrid_magnons}}
\end{center}\end{figure}

In recent years, hybrid quantum systems based on collective spin excitations in ferromagnetic crystals have become a promising platform for novel quantum technologies. Indeed, the quanta of these collective excitations, called magnons, can interact coherently with microwave and optical photons, as well as with phonons through magnetic dipole~\cite{Huebl2013,Tabuchi2014,Goryachev2014a,Zhang2014c,Tabuchi2016}, magneto-optical~\cite{Hisatomi2016,Osada2016,Zhang2016,Haigh2016,Kusminskiy2016,Sharma2017,Osada2017,Osada2018}, and magnetostrictive interactions~\cite{Zhang2015,Li2018}, respectively. As schematically shown in Fig.~\ref{fig:hybrid_magnons}, this diverse set of interactions allows one to further engineer coherent interactions between drastically different physical systems. Of particular interest is the architecture of quantum magnonics~\cite{Tabuchi2015,Tabuchi2016}, where the strong and coherent effective interaction between magnetostatic modes and superconducting quantum circuits enables one to reach the quantum regime of magnonics to resolve single magnons~\cite{Lachance-Quirion2017}, for example. More generally, hybrid quantum systems based on magnonics offer opportunities to develop novel quantum technologies such as microwave-to-optical quantum transducers~\cite{Hisatomi2016} for quantum information processing and quantum-enhanced detection of magnons for applications such as magnon spintronics~\cite{Karenowska2015} and dark matter searches~\cite{Barbieri1989}.

The theoretical background and experimental landscape of the field of cavity electromagnonics, in which the magnetic dipole coupling of magnetostatic modes to microwave-frequency cavity modes is investigated, are first discussed in Section~\ref{sec:cavity_magnonics}. Section~\ref{sec:quantum_magnonics} discusses the possibility of entering the quantum regime of magnonics by engineering an effective coherent interaction between magnetostatic modes and microwave-frequency superconducting circuits. Section~\ref{sec:optomagnonics} then discusses the fields of optomagnonics and cavity optomagnonics, which are both key platforms for future applications of hybrid quantum systems based on magnonics in the optical domain, such as microwave-to-optical quantum transducers. Section~\ref{sec:magnomechanics} discusses the relatively new subfield of cavity magnomechanics which combines cavity electromagnonics and the coupling between mechanical and magnetostatic modes through magnetostrictive forces. Finally, Section~\ref{sec:outlook} provides perspectives and outlook, as well as possible future applications, for hybrid quantum systems based on magnonics, with a particular focus on quantum magnonics. For clarity, Table~\ref{Table} summarizes the definitions of the excitations, bosonic operators, frequencies, and decay rates of the modes of the different physical systems considered in this review.

\begin{table*}
\begin{tabular}{l|c|c|c|l|l}
\hline 
Physical system & Mode & Excitation & Operators & Frequency & Decay rate\\
\hhline{======}
Microwave cavity & Microwave cavity mode & Microwave photon & $\hat a$, $\hat a^\dagger$ & $\omega_\mathrm{c}/2\pi\sim10$~GHz & $\kappa_\mathrm{c}/2\pi\sim1$~MHz\\
Optical cavity & Optical cavity mode & Optical photon & $\hat b$, $\hat b^\dagger$ & $\omega_\mathrm{o}/2\pi\sim200$~THz & $\kappa_\mathrm{o}/2\pi\sim1$~GHz\\
Ferromagnetic crystal & Magnetostatic mode & Magnon & $\hat c$, $\hat c^\dagger$ & $\omega_\mathrm{m}/2\pi\sim10$~GHz & $\gamma_\mathrm{m}/2\pi\sim1$~MHz\\
Ferromagnetic crystal & Deformation mode & Phonon & $\hat d$, $\hat d^\dagger$ & $\omega_\mathrm{d}/2\pi\sim10$~MHz & $\gamma_\mathrm{d}/2\pi\sim1$~kHz\\
Superconducting qubit & First qubit transition & Excited state & $\hat q$, $\hat q^\dagger$ & $\omega_\mathrm{q}/2\pi\sim10$~GHz & $\gamma_\mathrm{q}/2\pi\sim0.1$~MHz\\
\hline
\end{tabular}
\caption{\textbf{Definitions}. Excitations, bosonic operators, and typical values for the frequencies and decay rates of the modes of the different physical systems considered.
\label{Table}}
\end{table*}

\section{Cavity electromagnonics}
\label{sec:cavity_magnonics}

The interaction between collective degrees of freedom of solid-state systems and electromagnetic fields of cavities serves as the basis of circuit quantum electrodynamics~\cite{Blais2004,Wallraff2004a} and cavity optomechanics~\cite{Aspelmeyer2014} through electric dipole and radiation-pressure interactions, respectively. More recently, the strong and coherent magnetic dipole interaction between collective spin excitations in ensembles of spins and microwave cavities has been considered for both paramagnetic~\cite{Wesenberg2009,Imamoglu2009,Kubo2010,Schuster2010,Bushev2010,Bushev2011a,Chiorescu2010,Zhu2011,Amsuss2011,Abe2011,Kurizki2015} and ferromagnetic spin ensembles~~\cite{Huebl2013,Tabuchi2014,Goryachev2014a,Zhang2014c,Tabuchi2016}. Applications of such hybrid systems for quantum technologies include storage and retrieval of microwave fields~\cite{Julsgaard2013a,Grezes2015a}, quantum transduction~\cite{Williamson2014,OBrien2014,Hisatomi2016}, and quantum sensing~\cite{Degen2017,Bienfait2015,Bienfait2017}.

In this Section, the field of cavity electromagnonics, which makes use of the magnetic dipole interaction between magnetostatic modes in a ferromagnetic spin ensemble and the modes of a microwave cavity, is introduced. Cavity electromagnonics serves as the basis of almost all hybrid systems based on magnonics. Indeed, the magnetic dipole interaction between magnetostatic modes and microwave cavity modes is used in both quantum magnonics and cavity magnomechanics to engineer interactions with modes of physically distinct systems [Fig.~\ref{fig:hybrid_magnons}].

\subsection{Magnetostatic modes}

\subsubsection{Ferromagnetic spin ensemble}
We first start by describing an ensemble of $N$ spins $s$ of operator $\mathbf{\hat S}_i$ with nearest-neighbor ferromagnetic exchange interaction $J>0$ in an external magnetic field $\mathbf{B}_0$. The Hamiltonian of such a spin ensemble is given by
\begin{align}
\mathcal{\hat H}_\mathrm{s}=g^*\mu_\mathrm{B}\sum_i^N\mathbf{B}_0\cdot\mathbf{\hat S}_i-2J\sum_{\langle i,j\rangle}\mathbf{\hat S}_i\cdot\mathbf{\hat S}_j,
\label{eq:spin_hamiltonian}
\end{align}
where $g^*$ is the $g$-factor and $\mu_\mathrm{B}$ is the Bohr magneton~\cite{Kittel2011,Stancil2009}. While the first term describes the Zeeman effect, the second term describes the ferromagnetic exchange interaction between neighboring spins. Note that the second term is negligible for a paramagnetic spin ensemble as $J=0$.

The Hamiltonian of Eq.~\eqref{eq:spin_hamiltonian} can be written in terms of bosonic operators $\hat c_\mathbf{k}^\dagger$ and $\hat c_\mathbf{k}$ to get a picture based on spin waves described by a dispersion relation $\omega_\mathrm{m}(\mathbf{k})$
\begin{align}
\mathcal{\hat H}_\mathrm{m}/\hbar=\sum_\mathbf{k}\omega_\mathrm{m}(\mathbf{k})\hat c_\mathbf{k}^\dagger\hat c_\mathbf{k},
\label{eq:spin_waves}
\end{align}
where $\mathbf{k}$ is the wavevector of the spin-wave mode~\cite{Kittel2011,Stancil2009}. For example, for an infinite cubic lattice with a lattice parameter $a_0$, the dispersion relation in the long-wavelength limit is given by
\begin{align}
\hbar\omega_\mathrm{m}(k)\approx g^*\mu_\mathrm{B}B_0+2sJa_0^2k^2,
\end{align}
where $B_0=\left|\mathbf{B}_0\right|$ is the amplitude of the external magnetic field and $k=\left|\mathbf{k}\right|$ is the momentum. While the first term again describes the standard Zeeman effect, the second term shows that the exchange interaction lifts the degeneracy between modes with different momenta $k$.

\subsubsection{Magnetostatic limit}
In the static limit, corresponding to $ka_0\ll1$, the spin-wave modes correspond to magnetostatic modes, where the long-range dipole-dipole interactions between spins is dominant over the short-range exchange interactions~\cite{Stancil2009}. In that case, the Hamiltonian of Eq.~\eqref{eq:spin_waves} can be written as
\begin{align}
\mathcal{\hat H}_\mathrm{m}/\hbar=\sum_n\omega_\mathrm{m}^{(n)}\hat c_n^\dagger\hat c_n,
\label{eq:H_m}
\end{align}
where $n$ labels the magnetostatic mode of frequency $\omega_\mathrm{m}^{(n)}$ and with creation and annilation operations $\hat c_n^{\dagger}$ and $\hat c_n$, respectively. The geometry of the ferromagnetic crystal is important for determining the frequencies $\omega_\mathrm{m}^{(n)}$ of the modes. For example, in a sphere, the magnetostatic modes correspond to the Walker modes~\cite{Walker1957,Walker1958}. The simplest Walker mode is the uniform magnetostatic mode, or Kittel mode~\cite{Kittel1948}. In that case, $\hbar\omega_\mathrm{m}=g^*\mu_\mathrm{B}B_0$, where the mode index $n$ is omitted for simplicity. Therefore, for the Kittel mode, the energy $\hbar\omega_\mathrm{m}$ necessary to excite a single magnon corresponds to the energy necessary to excite a single spin $1/2$ in the same external magnetic field $B_0$. However, in stark constrast to a single spin $1/2$, the Kittel mode corresponds to a harmonic oscillator with creation and annihilation operators $\hat c^\dagger$ and $\hat c$, respectively. Indeed, as long as the number of magnons excited in the magnetostatic mode is much smaller than the number of spins $N$, the system is well approximated by a harmonic oscillator~\cite{Kittel2011}.

\subsection{Magnetic dipole interaction with a microwave cavity mode: theory}

\subsubsection{Microwave cavity mode}
\begin{figure}\begin{center}
\includegraphics[scale=1]{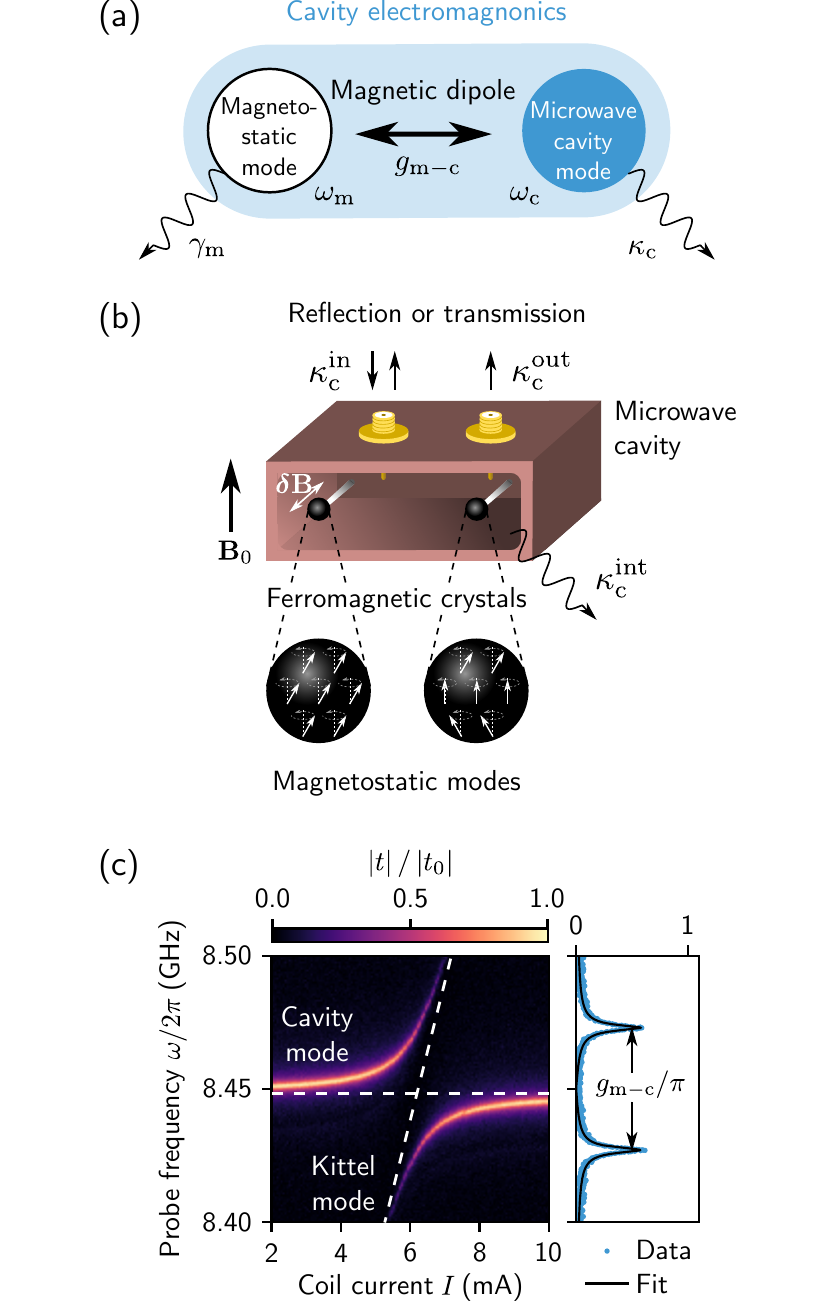}
\caption{\textbf{Cavity electromagnonics.}
(a)~Schematic diagram of the coupling between a magnetostatic mode and a microwave cavity mode of frequencies $\omega_\mathrm{m}$ and $\omega_\mathrm{c}$, respectively. The magnetic dipole interaction leads to a coupling strength $g_\mathrm{m-c}$ between these two modes. The magnetostatic and microwave cavity modes have linewidths given by $\gamma_\mathrm{m}$ and $\kappa_\mathrm{c}$, respectively. (b)~Schematic representation of a general hybrid system for cavity electromagnonics. The vacuum fluctuations $\boldsymbol{\delta}\mathbf{B}$ of the microwave magnetic field of a cavity mode overlaps with one or more ferromagnetic crystals. An external magnetic field $\mathbf{B}_0$ is applied either uniformly or locally to each ferromagnetic spin ensemble. Depending on the uniformity of the microwave magnetic field of the cavity mode, different magnetostatic modes can be coupled. The microwave cavity can be probed either by transmission or reflection through coupling rates $\kappa_\mathrm{c}^\mathrm{in}$ and $\kappa_\mathrm{c}^\mathrm{out}$ to input and output ports. The internal loss of the microwave cavity mode is given by $\kappa_\mathrm{c}^\mathrm{int}$. (c)~Amplitude of the transmission coefficient $\left|t\right|$ measured as a function of the probe frequency $\omega$ and the current $I$ in a coil that controls the amplitude $B_0$ of the static magnetic field. The amplitude of the transmission coefficient is normalized by its amplitude far from the avoided crossing, $\left|t_0\right|$. The clear avoided crossing indicates the strong and coherent coupling between the Kittel mode of a sphere of yttrium iron garnet (YIG) and a mode of a three-dimensional microwave cavity. The coupling strength $g_\mathrm{m-c}/2\pi=22.9$~MHz is determined from the spectrum measured when the two modes are resonant (right panel). Dashed lines are guides for the eye.
\label{fig:cavity_magnonics}}
\end{center}\end{figure}

The magnetic dipole interaction between a magnetostatic mode and a microwave cavity mode, depicted in Fig.~\ref{fig:cavity_magnonics}(a), is considered here. The Hamiltonian of a cavity mode of frequency $\omega_\mathrm{c}$ is given by
\begin{align}
\mathcal{\hat H}_\mathrm{c}/\hbar=\omega_\mathrm{c}\hat a^\dagger\hat a,
\label{eq:H_c}
\end{align}
where $\hat a^\dagger$ and $\hat a$ are the creation and annihilation operators of a microwave photon in the mode. It is worth noting that both magnetostatic modes and microwave cavity modes are linear systems described as quantum harmonic oscillators. Notably, as will be later discussed, this means that tools developed for quantum optics can readily be applied to magnonics in the quantum regime.

\subsubsection{Coupling to magnetostatic modes}
The magnetic dipole interaction of magnetostatic modes with microwave-frequency cavity modes is achieved by placing a ferromagnetic crystal so that it overlaps with the microwave magnetic field of the cavity mode~\cite{Imamoglu2009}. This magnetic dipole interaction is described by the Hamiltonian
\begin{align}
\mathcal{\hat H}_\mathrm{s-c}=g^*\mu_\mathrm{B}\left(\hat a+\hat a^\dagger\right)\sum_i^N\boldsymbol\delta\mathbf{B}(\mathbf{r}_i)\cdot\mathbf{\hat S}_i,
\label{eq:magnetic_dipole_interaction}
\end{align}
where $\mathbf{B}_1(\mathbf{r}_i)=\boldsymbol\delta\mathbf{B}(\mathbf{r}_i)\left(\hat a+\hat a^\dagger\right)$ is the microwave cavity magnetic field at position $\mathbf{r}_i$ of spin $i$, and $\boldsymbol\delta\mathbf{B}(\mathbf{r}_i)$ is the amplitude of the corresponding vacuum fluctuations. Notably, this interaction Hamiltonian is the same for a paramagnetic spin ensemble~\cite{Imamoglu2009,Wesenberg2009}.

Considering a linearly-polarized microwave cavity magnetic field and applying the rotating wave approximation, the Hamiltonian of Eq.~\eqref{eq:magnetic_dipole_interaction} can be expressed in terms of the magnetostatic modes in the ferromagnetic spin ensemble as
\begin{align}
\mathcal{\hat H}_\mathrm{m-c}/\hbar=\sum_n g_\mathrm{m-c}^{(n)}\left(\hat a^\dagger\hat c_n+\hat a\hat c_n^\dagger\right),
\end{align}
where $g_\mathrm{m-c}^{(n)}$ is the coupling strength between the magnetostatic mode $n$ and the microwave cavity mode~\cite{Tabuchi2016}. The magnetic dipole coupling strength is given by
\begin{align}
\hbar g_\mathrm{m-c}^{(n)}=\frac{1}{2}\sqrt{2s}g^*\mu_\mathrm{B}\int_V\mathrm{d}\mathbf{r}\ \boldsymbol{\delta}\mathbf{B}(\mathbf{r})\cdot\mathbf{s}_n(\mathbf{r}),
\label{eq:general_cavity_magnon_coupling}
\end{align}
where the interaction is integrated throughout the volume $V$ of the ferromagnetic crystal, and $\mathbf{s}_n(\mathbf{r})$ is the orthonormal mode function describing the spatial profile of the amplitude and phase for the magnetostatic mode $n$.

\subsubsection{Coupling to the Kittel mode}
If the microwave magnetic field $\boldsymbol{\delta}\mathbf{B}(\mathbf{r})$ is uniform throughout the ferromagnetic crystal, the magnetic dipole coupling vanishes except for the uniform magnetostatic mode, i.e., the Kittel mode~\cite{Tabuchi2016}. As will be later discussed, coupling between higher-order magnetostatic modes and a microwave cavity mode therefore requires a non-uniform cavity magnetic field. Considering only the Kittel mode, the Hamiltonian of the interaction is given by
\begin{align}
\mathcal{\hat H}_\mathrm{m-c}/\hbar=g_\mathrm{m-c}\left(\hat a^\dagger\hat c+\hat a\hat c^\dagger\right),
\label{eq:H_m-c}
\end{align}
where the coupling strength $g_\mathrm{m-c}$ between the Kittel mode and the microwave cavity mode is given by
\begin{align}
\hbar g_\mathrm{m-c}=\frac{1}{2}\sqrt{2s}g^*\mu_\mathrm{B}\delta B\sqrt{N},
\label{eq:cavity_magnon_coupling_strength}
\end{align}
where $\delta B\equiv\left|\boldsymbol{\delta}\mathbf{B}\right|$ is the amplitude of the vacuum fluctuations of the microwave cavity magnetic field in the ferromagnetic crystal transverse to the external magnetic field $\mathbf{B}_0$. As in a paramagnetic spin ensemble, a $\sqrt{N}$ collective enhancement of the coupling strength between a single spin and the microwave cavity mode is realized by using an ensemble of $N$ spins~\cite{Wesenberg2009,Imamoglu2009,Schuster2010,Kubo2010,Abe2011}. As depicted in Fig.~\ref{fig:cavity_magnonics}(a), the resulting coupling strength $g_\mathrm{m-c}$ between the Kittel mode and the microwave cavity mode has to be compared with the linewidth of the Kittel mode, $\gamma_\mathrm{m}$, and the linewidth of the cavity mode, $\kappa_\mathrm{c}$. When $g_\mathrm{m-c}\gg\gamma_\mathrm{m},\kappa_\mathrm{c}$, the hybrid system enters the strong coupling regime.

\subsection{Magnetic dipole interaction with a microwave cavity mode: experiments}

\subsubsection{Strong coupling regime}
Magnetic dipole coupling strengths of tens of MHz between microwave cavity modes and paramagnetic spin ensembles with spin densities of $10^{16}-10^{18}~\mu_\mathrm{B}\ \mathrm{cm}^{-3}$ have been previously demonstrated~\cite{Kubo2010,Schuster2010,Abe2011,Amsuss2011,Bushev2011a}. As ferromagnetic materials have much larger spin densities of about $10^{22}~\mu_\mathrm{B}\ \mathrm{cm}^{-3}$, larger coupling strengths between the magnetostatic modes of a ferromagnetic crystal and microwave cavity modes are expected~\cite{Huebl2013,Tabuchi2014}. By combining these large coupling strengths with the linewidths of magnetostatic modes, which are in the MHz range even at room temperature, the strong coupling regime of cavity electromagnonics can be readily reached. Furthermore, as opposed to paramagnetic spin ensembles, the ferromagnetic exchange interaction between neighboring spins offers a rigidity to the magnetostatic modes that favors long-wavelength excitations.

The first demonstration of the strong coupling regime in cavity electromagnonics was achieved in an experiment which used a rectangular gallium-doped yttrium iron garnet (YIG) crystal placed directly on top of a superconducting resonator~\cite{Huebl2013}, analogous to pioneering experiments with paramagnetic spin ensembles~\cite{Kubo2010,Schuster2010,Abe2011,Amsuss2011,Bushev2011a}. However, due to the inherent inhomogeneity of the microwave magnetic field of the superconducting resonator and the large magnon linewidth due to the gallium doping, the normal-mode splitting was not observed despite the hybrid system reaching the strong coupling regime.

These issues were addressed in subsequent experiments by using three-dimensional microwave cavities and spherical crystals of undoped YIG [Fig.~\ref{fig:cavity_magnonics}(b)], enabling the observation of the normal-mode splitting at both cryogenic temperatures~\cite{Tabuchi2014,Goryachev2014a} and at room temperature~\cite{Zhang2014c}. Indeed, as both systems are linear and are coupled through a linear beam splitter-like interaction, the underlying physics is mostly classical. Therefore, thermal populations of the different modes do not significantly affect the dynamics of the coupled hybrid system and the normal-mode splitting can be observed even at room temperature~\cite{Gurevich1996}.

\subsubsection{Example of the strong coupling regime}
As an illustrative example, Fig.~\ref{fig:cavity_magnonics}(c) shows a measurement of the normal-mode splitting between the Kittel mode of a YIG sphere and a mode of a three-dimensional microwave cavity at millikelvin temperatures. The microwave cavity is a three-dimensional cavity made out of copper with dimensions of $24\times3\times53$~mm$^3$. The lowest frequency modes are the transverse electric (TE) modes TE$_{10p}$ of frequencies $\omega_{10p}$ and linewidths $\kappa_{10p}$. For example, the TE$_{102}$ mode has a frequency $\omega_\mathrm{102}/2\pi=8.41$~GHz and a linewidth $\kappa_\mathrm{102}/2\pi=2.1$~MHz. The two ports of the microwave cavity enable one to probe the cavity modes either in reflection or transmission.

A YIG sphere with a diameter of $0.5$~mm is placed inside the microwave cavity near the antinode of the microwave magnetic field of the TE$_{102}$ mode. The uniformity of the microwave cavity magnetic field $\boldsymbol{\delta}\mathbf{B}(\mathbf{r})$ throughout the sphere is better than $1\%$ for the TE$_{102}$ mode, highly favoring the magnetic dipole coupling to the Kittel mode compared to higher-order modes. A static magnetic field $\mathbf{B}_0$ is applied along the $\langle100\rangle$ crystal axis and perpendicular to $\boldsymbol{\delta}\mathbf{B}$. The amplitude $B_0$ of the static field sets the magnon frequency $\omega_\mathrm{m}=g^*\mu_\mathrm{B}B_0/\hbar$ and is created by a combination of a pair of permanent magnets, a yoke made out of pure iron, and a $10^4$-turn superconducting coil. The disk-shaped neodymium permanent magnets produce a uniform static magnetic field $B_0\approx0.29$~T at the YIG sphere inside the microwave cavity. The coil enables one to change \textit{in situ} the magnon frequency by varying the current $I$ in the coil with a proportionality constant of $\sim1.7$~mT/mA.

The device is placed inside a dilution refrigerator with a base temperature of a few tens of millikelvins, leading to a thermal occupancy much smaller than unity of every relevant mode of the hybrid system. Furthermore, the microwave line connected to the input port of the microwave cavity is highly attenuated to prevent further thermal excitations from stages of the dilution refrigerator at higher temperatures. The output port is connected to cryogenic and room-temperature low-noise microwave amplifiers. The device is isolated from the thermal noise of the amplifiers by cryogenic circulators and isolators. More details on the experimental setup can be found in Refs.~\citenum{Tabuchi2015,Tabuchi2016}, and \citenum{Lachance-Quirion2017}.

The normalized amplitude of the transmission coefficient, $\left|t\right|/\left|t_0\right|$, is shown in Fig.~\ref{fig:cavity_magnonics}(c) as a function of the current $I$ in the coil, effectively changing the magnon frequency $\omega_\mathrm{m}$. The microwave cavity is probed at a power corresponding to, on average, much less than a single photon populating the cavity mode. Due to the large magnetic dipole interaction, when the Kittel mode is close to resonance with the TE$_{102}$ cavity mode, the two modes hybridize, leading to a normal-mode splitting; the hallmark of the strong coupling regime of cavity electromagnonics~\cite{Huebl2013,Tabuchi2014,Goryachev2014a,Zhang2014c}. Indeed, the coupling strength $g_\mathrm{m-c}/2\pi=22.9$~MHz, extracted from the spectrum measured when the hybrid system is fully hybridized, is much larger than the linewidths of the Kittel and microwave cavity modes, $\gamma_\mathrm{m}/2\pi=1.4$~MHz and $\kappa_\mathrm{c}/2\pi=2.1$~MHz, respectively. The magnon linewidth is determined from the linewidth $\left(\gamma_\mathrm{m}+\kappa_\mathrm{c}\right)/2$ when $\omega_\mathrm{m}=\omega_\mathrm{c}$ and from a measurement of the microwave cavity linewidth $\kappa_\mathrm{c}$ far from the avoided crossing.

\subsection{Further experimental work}

\subsubsection{Materials}
Early works on cavity electromagnonics have mainly focused on using YIG as the ferromagnetic material~\cite{Huebl2013,Tabuchi2014,Goryachev2014a,Zhang2014c,Bai2015}. These experiments prompted the study of the properties of ferromagnetic materials at millikelvin temperatures, which did result in novel observations. Notably, in Ref.~\citenum{Tabuchi2014}, the temperature dependence of the linewidth $\gamma_\mathrm{m}$ of the Kittel mode was investigated from a few Kelvins down to about $10$~mK. When decreasing the temperature down to about $1$~K, the linewidth decreases monotonically as expected from slow relaxation due to impurity ions and losses due to magnon-phonon scattering~\cite{Teale1962,Sparks1961}. However, for temperatures below $1$~K, the magnon linewidth increases again and saturates at $\gamma_\mathrm{m}/2\pi\approx1$~MHz at a temperature of about $100$~mK. This behavior can be understood by the presence of a bath of two-level systems (TLSs) resonant with the Kittel mode, similar to losses from TLSs in superconducting resonators~\cite{Muller2017}. Indeed, for a temperature $T\ll\hbar\omega_\mathrm{m}/k_\mathrm{B}$, the bath of resonant TLSs acts as a new decay channel for magnons. For $k_\mathrm{B}T\gg\hbar\omega_\mathrm{m}$, however, the TLSs are saturated and therefore do not contribute to magnon decay. The observed temperature dependence of the magnon linewidth is well reproduced by this model~\cite{Tabuchi2014}. The origin of the bath of TLSs however remains to be clearly identified and is an important open question for hybrid quantum systems based on magnonics~\cite{Tabuchi2016}. While losses attributed to TLSs make up a large fraction of the losses at millikelvin temperatures, the remaining linewidth can be ascribed to elastic magnon-magnon scattering at the surface of the ferromagnetic crystal~\cite{Sparks1961,Tabuchi2014}. Fortunately, this contribution to the magnon linewidth can probably be improved by decreasing the surface roughness of the ferromagnetic sample.

Subsequent works in cavity electromagnonics have also been performed on other materials. For example, the strong coupling between magnetostatic modes in lithium ferrite and a microwave cavity has been demonstrated~\cite{Goryachev2017}. Interestingly, the softening of the magnetostatic modes at low magnetic fields enables the hybrid system to reach a regime where the magnon frequency is both first-order and second-order insensitive to the amplitude of the external magnetic field, in analogy to clock transitions in atomic and spin systems~\cite{Bollinger1985,Wolfowicz2013b}. Furthermore, the multiferroic chiral magnetic insulator Cu$_2$OSeO$_3$ has also been investigated in the strong coupling regime of cavity electromagnonics~\cite{Abdurakhimov2018}. Notably, the phase transition of the material from paramagnetic to ferromagnetic is observed through a change in the coupling strength between the collective spin excitations in the material and the microwave cavity mode. These early results hint that cavity electromagnonics could be used as a novel tool to characterize materials with a ferromagnetic order.

\subsubsection{Microwave cavity geometries}
Many different geometries of microwave cavities have been investigated in cavity electromagnonics. In addition to the three-dimensional microwave cavities previously discussed [Fig.~\ref{fig:cavity_magnonics}(b)], superconducting coplanar waveguide resonators~\cite{Huebl2013,Morris2017}, coaxial-like cavities~\cite{Haigh2015,Bourhill2016} and lumped-element cavities~\cite{Goryachev2014a,Goryachev2017a,McKenzie-Sell2019,Yang2019,Bhoi2019} have been used to reach the strong coupling regime. In particular, lumped-element cavities, also used in experiments with paramagnetic spin ensembles~\cite{Bienfait2015,Bienfait2016,Angerer2016a,Astner2018,Angerer2018,Ball2018}, enable one to achieve larger filling factors $\eta$ of the microwave cavity magnetic field within the volume of the ferromagnetic crystal compared to a three-dimensional cavity of the same frequency, leading to a larger magnetic dipole coupling strength $g_\mathrm{m-c}\propto\sqrt{\eta}$~\cite{McKenzie-Sell2019}. Furthermore, antinodes of the microwave cavity magnetic and electric fields can be more easily spatially separated in a lumped-element cavity as they are respectively located at the inductive and capacitive elements. This is a great advantage for adding elements, such as superconducting qubits, that couple to the electric field of the microwave cavity through an electric dipole interaction and that are adversely affected by the external magnetic field applied to the ferromagnetic crystal~\cite{Tabuchi2015}. Finally, as lumped-element cavities have an effective dimensionality of zero, smaller cavities can be designed while keeping the frequencies of the modes at about $10$~GHz~\cite{McKenzie-Sell2019}. As discussed in Section~\ref{sec:optomagnonics}, this could be particularly advantageous for cavity optomagnonics, where the optomagnonic coupling is increased in smaller ferromagnetic samples~\cite{Kusminskiy2016}.

\subsubsection{Multiple modes and multiple ferromagnetic crystals}
Thus far, the interaction between a single microwave cavity mode and a single magnetostatic mode in a ferromagnetic crystal has been discussed. The theory presented earlier can readily be adapted to consider the multimode nature of both the ferromagnetic crystal and most types of cavities simply by summing the system and interaction Hamiltonians over all the relevent modes. Furthermore, devices incorporating multiple ferromagnetic crystals within the mode volume of a single microwave cavity mode have been investigated~\cite{Zhang2015c,Lambert2016,Zhang2018,ZareRameshti2018}. Notably, the coupling between the collective modes of strongly coupled Kittel modes in up to 8 spheres to a common microwave cavity mode has enabled one to realize a gradient memory in dark collective modes~\cite{Zhang2015c}. Furthermore, an effective coupling between the Kittel modes of two spatially separated spheres has been achieved by using the commonly coupled cavity mode as a coupling bus~\cite{Lambert2016}. These results show that coupling magnetostatic modes to microwave cavity modes provides a new way to couple distinct magnetic systems, such as ferromagnetic and antiferromagnetic samples~\cite{Johansen2018}.

\subsubsection{Higher-order magnetostatic modes}
In ferromagnetic spheres, the spatial uniformity of the Kittel mode favors coupling to the microwave cavity modes compared to higher-order magnetostatic modes when the cavity mode is uniform throughout the ferromagnetic sample. This enables one to greatly simplify the description of the hybrid system by having to only consider a single magnetostatic mode strongly coupled to microwave cavity modes~\cite{Tabuchi2014,Zhang2014c}. Even then, in most experiments, hints of a finite coupling to higher-order magnetostatic modes are observed in the measurement of the avoided crossing between the Kittel mode and a microwave cavity mode~\cite{Zhang2014c,Tabuchi2015}. Experiments voluntarity having a nonuniform microwave magnetic field throughout the volume of the ferromagnetic sphere have been performed to investigate the coupling to higher-order magnetostatic modes~\cite{Goryachev2014a,Morris2017}. This is of particular interest for cavity optomagnonics, discussed in Section~\ref{sec:optomagnonics}, as the optomagnonic coupling is expected to be larger for higher-order modes~\cite{Kusminskiy2016,Sharma2017,Haigh2018}. To this end, a tomography technique has been recently devised to experimentally characterize magnetostatic modes in a ferromagnetic sphere~\cite{Gloppe2018}.

\subsubsection{Dissipative coupling}
Up to now, the coherent coupling between a magnetostatic mode and a microwave cavity mode due to the magnetic dipole interaction has been discussed. Recently, it has been shown that a dissipative coupling due to the Lenz effect is also possible in cavity electromagnonics~\cite{Harder2018,Yang2019,Bhoi2019}. For example, when the ferromagnetic crystal is placed at the node of the magnetic field of a microwave cavity mode, the coherent coupling vanishes [Eq.~\eqref{eq:cavity_magnon_coupling_strength}]. However, the cavity Lenz effect induces a microwave current in the cavity when magnons are excited, which in turn impedes the excitation of magnons~\cite{Harder2018}. This leads to a level attraction when the hybrid system is on resonance, in stark contrast to the usual avoided crossing observed for a coherent coupling [Fig.~\ref{fig:cavity_magnonics}(c)]~\cite{Gloppe2014}. As discussed in Ref.~\citenum{Harder2018}, this hints that great care needs to be taken when evaluating the coherent coupling strength $g_\mathrm{m-c}$ from the avoided crossing as a finite dissipative coupling will affect the experimentally observed normal-mode splitting. In the presence of both coherent and dissipative couplings, the interaction Hamiltonian of Eq.~\eqref{eq:H_m-c} between a magnetostatic mode and a microwave cavity mode can be generalized as
\begin{align}
\mathcal{\hat H}_\mathrm{m-c}/\hbar=g_\mathrm{m-c}\left(\hat a^\dagger\hat c+e^{i\Phi}\hat a\hat c^\dagger\right),
\end{align}
where the coupling phase $\Phi$ describes the competing coherent and dissipative couplings~\cite{Harder2018}. The Hamiltonian of Eq.~\eqref{eq:H_m-c}, describing only the coherent coupling, corresponds to the case $\Phi=0$.

\section{Quantum magnonics}
\label{sec:quantum_magnonics}

A promising approach to observe quantum effects in magnonics is to consider a nonlinear system interacting through a linear, beam splitter-like interaction with the magnetostatic modes. Such nonlinear quantum systems can be implemented in superconducting circuits where the Josephson effect provides the nonlinearity necessary to use these circuits as qubits~\cite{Nakamura1999,Devoret2013}. Furthermore, superconducting qubits can interact strongly with microwave cavity modes through a Jaynes-Cummings-like electric dipole interaction in the cQED paradigm~\cite{Blais2004,Wallraff2004a} in close relation to cavity quantum electrodynamics~\cite{Raimond2001,Haroche2006}. In quantum magnonics, this qubit-cavity interaction is combined with the beam-splitter-like magnetic dipole interaction between the microwave cavity modes and magnetostatic modes of a ferromagnetic crystal to provide the nonlinearity necessary to explore quantum effects in magnonics~\cite{Tabuchi2015,Tabuchi2016,Lachance-Quirion2017}. We note that a similar approach was investigated in paramagnetic spin ensembles~\cite{Kubo2011,Kubo2012a}.

\subsection{Coupling to a qubit: theory}

\subsubsection{Superconducting qubit}
\begin{figure}\begin{center}
\includegraphics[scale=1]{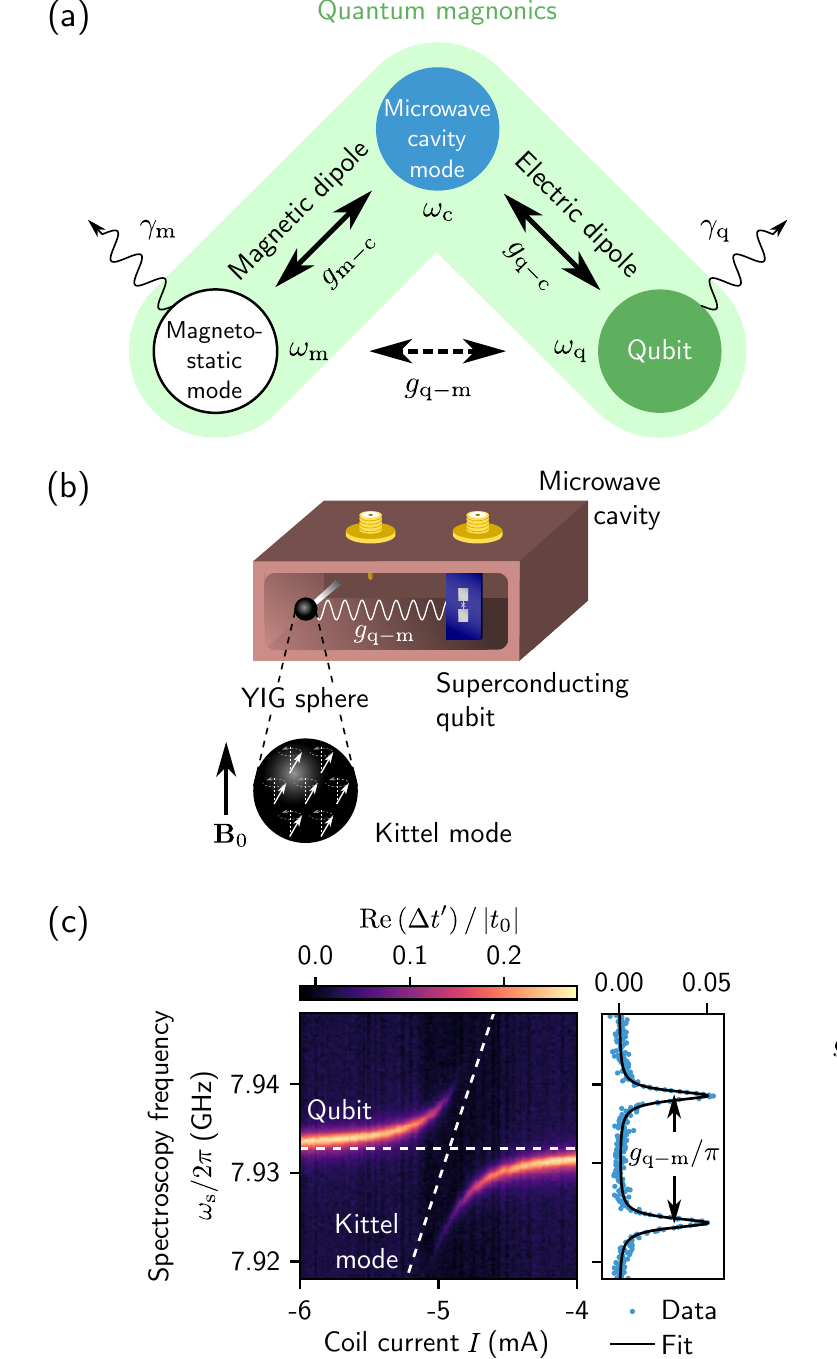}
\caption{\textbf{Quantum magnonics.}
(a)~Schematic diagram of the coupling between a magnetostatic mode and a superconducting qubit of frequencies $\omega_\mathrm{m}$ and $\omega_\mathrm{q}$, respectively. Both systems are coupled to a common microwave cavity mode of frequency $\omega_\mathrm{c}$ through magnetic and electric dipole interactions of coupling strengths $g_\mathrm{m-c}$ and $g_\mathrm{q-c}$, respectively. This leads to a second-order effective coupling between the magnetostatic mode and the superconducting qubit of coupling strength $g_\mathrm{q-m}$. The strong coupling regime is reached when $g_\mathrm{q-m}$ is larger than both magnon and qubit linewidths $\gamma_\mathrm{m}$ and $\gamma_\mathrm{q}$, respectively.
(b)~Schematic representation of the device used to demonstrate the strong coupling between the uniform magnetostatic mode, or Kittel mode, of a YIG sphere and a transmon-type superconducting qubit. The frequency of the magnons in the Kittel mode is tunable through the amplitude of the external magnetic field $\left|\mathbf{B}_0\right|$. Both systems are placed inside a multimode three-dimensional microwave cavity, leading to an effective qubit-magnon coupling.
(c)~Relative change of the transmission coefficient $\mathrm{Re}\left(\Delta t'\right)/\left|t_0\right|$ of the microwave cavity mode at a probe frequency $\omega\approx\omega_\mathrm{c}$ as a function of the spectroscopy frequency $\omega_\mathrm{s}\approx\omega_\mathrm{q}$ and the current $I$ in a coil changing the magnon frequency $\omega_\mathrm{m}$. The qubit-magnon coupling strength $g_\mathrm{q-m}/2\pi=7.2$ MHz, determined from the spectrum shown in the right panel, puts the hybrid system in the strong coupling regime. Dashed lines are guides for the eye.
\label{fig:quantum_magnonics}}
\end{center}\end{figure}

The hybrid system considered in quantum magnonics, depicted schematically in Fig.~\ref{fig:quantum_magnonics}(a), is composed of a ferromagnetic crystal, a microwave cavity, and a superconducting qubit. The superconducting qubit can be described as an anharmonic oscillator by the Hamiltonian
\begin{align}
\mathcal{\hat H}_\mathrm{q}/\hbar=\left(\omega_\mathrm{q}-\frac{\alpha}{2}\right)\hat q^\dagger\hat q+\frac{\alpha}{2}\left(\hat q^\dagger\hat q\right)^2,
\label{eq:H_q}
\end{align}
where $\omega_\mathrm{q}$ is the frequency of the transition between the ground state $|g\rangle$ and the first excited state $|e\rangle$, $\alpha$ is the anharmonicity, and $\hat q^\dagger$ and $\hat q$ are respectively the creation and annihilation operators for the qubit. The frequency of the transition between the first and second excited states is given by $\omega_\mathrm{q}+\alpha$. For the so-called transmon regime of a superconducting qubit, $\alpha$ is negative and sufficiently large to operate such anharmonic oscillators as qubits~\cite{Koch2007}. The coherence time $T_2^*$ of the qubit is related to its linewidth $\gamma_\mathrm{q}$ through $\gamma_\mathrm{q}=2/T_2^*$. For example, for a very modest coherence time $T_2^*=1~\mu$s, the qubit linewidth $\gamma_\mathrm{q}/2\pi=0.32$~MHz is already smaller than the linewidth $\gamma_\mathrm{m}/2\pi\sim1$~MHz of the Kittel mode at millikelvin temperatures~\cite{Tabuchi2014}.

\subsubsection{Description of the hybrid system}
The large electric dipole of superconducting qubits enables them to strongly couple to the electric field of the modes of planar superconducting resonators and three-dimensional microwave cavities~\cite{Wallraff2004a,Paik2011}. Under the rotating wave approximation, the qubit-cavity electric dipole interaction is described by the Hamiltonian
\begin{align}
\mathcal{\hat H}_\mathrm{q-c}/\hbar=g_\mathrm{q-c}\left(\hat q^\dagger\hat a+\hat q\hat a^\dagger\right),
\label{eq:H_q-c}
\end{align}
where $g_\mathrm{q-c}$ is the qubit-cavity coupling strength~\cite{Blais2004}. It is therefore possible to describe the hybrid system considered in quantum magnonics by the Hamiltonian
\begin{align}
\mathcal{\hat H}=\mathcal{\hat H}_\mathrm{m}+\mathcal{\hat H}_\mathrm{c}+\mathcal{\hat H}_\mathrm{q}+\mathcal{\hat H}_\mathrm{m-c}+\mathcal{\hat H}_\mathrm{q-c},
\label{eq:Total_hamiltonian_quantum_magnonics}
\end{align}
where the different terms are given by Eqs.~\eqref{eq:H_m}, \eqref{eq:H_c}, \eqref{eq:H_q}, \eqref{eq:H_m-c}, and \eqref{eq:H_q-c}, respectively.

\subsubsection{Effective coupling}
The individual interactions of both the magnetostatic mode and the qubit to common microwave cavity modes enable one to engineer an effective interaction between these two very distinct macroscopic physical systems~\cite{Tabuchi2015,Tabuchi2016}. When the magnetostatic mode and the qubit are close to resonance and far detuned from the microwave cavity, so that $\left|\omega_\mathrm{q}-\omega_\mathrm{m}\right|\ll g_\mathrm{q-c},g_\mathrm{m-c}\ll\left|\omega_\mathrm{q}-\omega_\mathrm{c}\right|,\left|\omega_\mathrm{m}-\omega_\mathrm{c}\right|$, the effective qubit-magnon interaction is described by the Jaynes-Cummings-type Hamiltonian
\begin{align}
\mathcal{\hat H}_\mathrm{q-m}/\hbar=g_\mathrm{q-m}\left(\hat q^\dagger\hat c+\hat q\hat c^\dagger\right),
\end{align}
where the microwave cavity modes have been adiabatically eliminated~\cite{Imamoglu2009,Tabuchi2015,Tabuchi2016}. With the magnetostatic mode and the qubit on resonance, such that $\omega_\mathrm{m}=\omega_\mathrm{q}\equiv\omega_\mathrm{q,m}$, the coupling strength $g_\mathrm{q-m}$ is given by
\begin{align}
g_\mathrm{q-m}\approx\sum_p\frac{g_\mathrm{m-c}^{(p)}g_\mathrm{q-c}^{(p)}}{\omega_\mathrm{q,m}-\omega_\mathrm{c}^{(p)}},
\label{eq:qubit-magnon_coupling_strength}
\end{align}
where $\omega_\mathrm{c}^{(p)}$ is the frequency of the microwave cavity mode of index $p$ with magnetic and electric dipole coupling strengths $g_\mathrm{m-c}^{(p)}$ and $g_\mathrm{q-c}^{(p)}$ to the magnetostatic mode and the qubit, respectively~\cite{Tabuchi2016}. The regime of strong qubit-magnon coupling is reached when $g_\mathrm{q-m}$ is much larger than the linewidths of both the magnetostatic mode and the qubit, $\gamma_\mathrm{m}$ and $\gamma_\mathrm{q}$, respectively. 

\subsection{Coupling to a qubit: experiment}

\subsubsection{Device}
Figure~\ref{fig:quantum_magnonics}(b) shows a schematic representation of the device used to demonstrate the strong coupling regime of quantum magnonics~\cite{Tabuchi2015}. The results mentioned here are with the same device as in Section~\ref{sec:cavity_magnonics} and Ref.~\citenum{Lachance-Quirion2017}. The transmon-type superconducting qubit is made out of two large-area aluminum antenna pads connected by a Josephson junction (Al/Al$_2$O$_3$/Al) fabricated on a silicon substrate. The bare frequency $\omega_\mathrm{q}/2\pi=7.97$~GHz and anharmonicity $\alpha/2\pi=-0.14$~GHz of the qubit are set by the Josephson energy $E_J/h\approx60$~GHz and the charging energy $E_C/h\approx0.14$~GHz~\cite{Koch2007}. The qubit is placed close to the antinode of the electric field of the TE$_{102}$ cavity mode and is separated by about $35$~mm from the YIG sphere. A magnetic shield made of aluminum and pure iron covers half of the microwave cavity to protect the qubit from the stray magnetic field of the magnetic circuit used to apply the external magnetic field.

\subsubsection{Experiment}
The effective coupling between the Kittel mode and the qubit is investigated by performing two-tone spectroscopy of the qubit. Indeed, the dispersive qubit-cavity interaction enables one to measure the qubit absorption spectrum by probing the change in reflection or transmission of one of the microwave cavity modes~\cite{Wallraff2005,Schuster2005,Gambetta2006}. While the qubit-magnon coupling is mainly mediated by the TE$_{102}$ cavity mode, the TE$_{103}$ cavity mode is used for reading out the qubit~\cite{Tabuchi2015,Lachance-Quirion2017}. Figure~\ref{fig:quantum_magnonics}(c) shows the measured qubit spectrum close to resonance with the Kittel mode. The clear avoided crossing indicates that the engineered effective interaction has reached the strong coupling regime. Indeed, the coupling strength $g_\mathrm{q-m}/2\pi=7.2$~MHz is much larger than the linewidths of the Kittel mode and the qubit.

While the coil current can be used to control \textit{in situ} the detuning between the qubit and the magnetostatic modes, the strength of the qubit-magnon coupling $g_\mathrm{q-m}$ is determined by the static detuning between $\omega_\mathrm{q}=\omega_\mathrm{m}$ and the microwave cavity modes [Eq.~\eqref{eq:qubit-magnon_coupling_strength}]. A dynamically tunable qubit-magnon interaction can be engineered by using a parametric drive at a frequency given by the average frequency of the detuned Kittel mode and qubit~\cite{Tabuchi2015}. While the resulting coupling strength is smaller than for the static case, the ability to activate the coupling on-demand on a nanosecond timescale without any modifications to the hardware of the hybrid system can be useful to transfer excitations between the Kittel mode and the qubit in a controllable way~\cite{Wallraff2007,Leek2009,Zakka-Bajjani2011,Flurin2014a,Tabuchi2015}.

\subsection{Dispersive coupling}

\subsubsection{Dispersive regime}
A dynamically tunable coupling or detuning can be used to prepare quantum states in bosonic modes~\cite{Law1996,Hofheinz2008,Hofheinz2009,Satzinger2018,Chu2018}. Alternatively, a strong and static dispersive coupling enables one to prepare and characterize quantum states in a harmonic oscillator~\cite{Haroche2006,Vlastakis2013a,Kirchmair2013} such as the Kittel mode. The qubit-magnon hybrid system enters the dispersive regime when the detuning between the qubit and the Kittel mode is much larger than the static coupling strength, i.e., $\left|\omega_\mathrm{q}-\omega_\mathrm{m}\right|\gg g_\mathrm{q-m}$. In analogy to the qubit-cavity dispersive regime in cavity and cQED~\cite{Haroche2006,Blais2004}, the Hamiltonian of the interaction between a magnetostatic mode and the superconducting qubit in the dispersive regime is given by
\begin{align}
\mathcal{\hat H}_\mathrm{q-m}^\mathrm{disp}/\hbar=2\chi_\mathrm{q-m}\hat q^\dagger\hat q\hat c^\dagger\hat c,
\label{eq:dispersive_interaction}
\end{align}
where $\chi_\mathrm{q-m}$ is the strength of the dispersive interaction. The dispersive coupling therefore shifts the qubit frequency by $2\chi_\mathrm{q-m}$ for every magnon in the Kittel mode, and, reciprocally, shifts the frequency of the Kittel mode by $2\chi_\mathrm{q-m}$ when the qubit is excited.

As the interaction between the Kittel mode and the qubit is a second-order process mediated by the couplings to the microwave cavity modes, the effective dispersive interaction is a fourth-order process, leading to the ordering $\left|\chi_\mathrm{q-m}\right|\ll g_\mathrm{q-m}\ll g_\mathrm{m-c},g_\mathrm{q-c}$. While an analytical expression of $\chi_\mathrm{q-m}$ is given in Ref.~\citenum{Tabuchi2015}, it can also be calculated directly by diagonalizing numerically the Hamiltonian of the hybrid system given by Eq.~\eqref{eq:Total_hamiltonian_quantum_magnonics}~\cite{Lachance-Quirion2017}.

\subsubsection{Strong dispersive regime}
\begin{figure}\begin{center}
\includegraphics[scale=1]{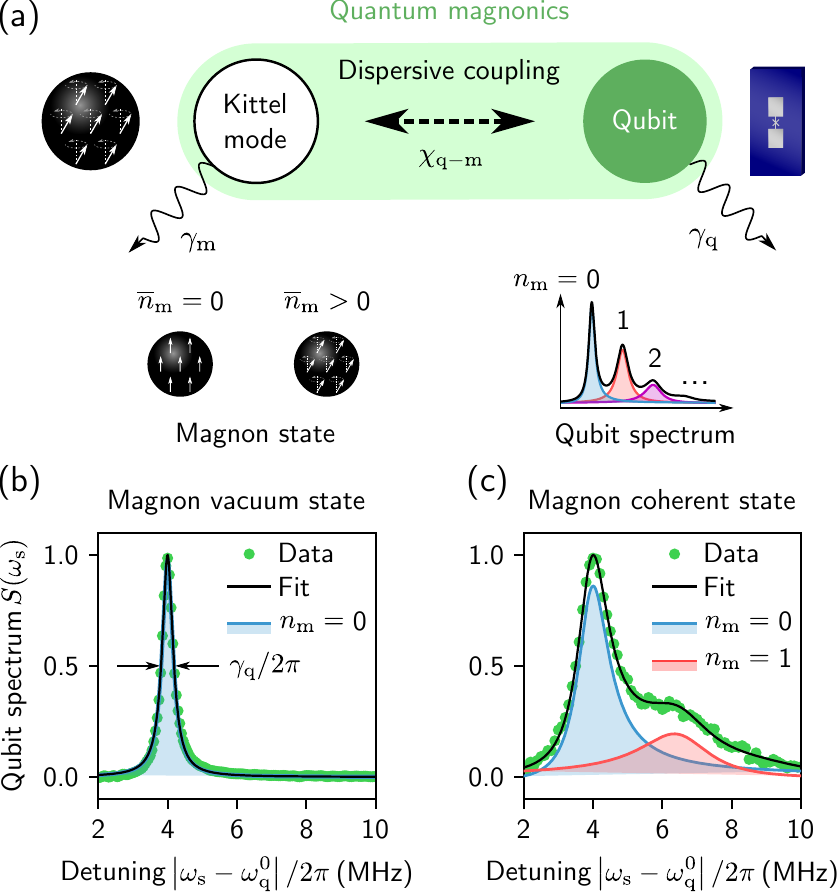}
\caption{\textbf{Resolving single magnons.}
(a)~Schematic diagram of the dispersive coupling between the Kittel mode and the superconducting qubit. The strong dispersive regime corresponds to a shift per magnon, $2\left|\chi_\mathrm{q-m}\right|$, larger than both $\gamma_\mathrm{m}$ and $\gamma_\mathrm{q}$, respectively. In this regime, when the Kittel mode is coherently driven to a nonzero magnon population $\overline{n}_\mathrm{m}>0$, the qubit spectrum splits into multiple peaks, each corresponding to a given magnon number state $|n_\mathrm{m}\rangle$. (b, c)~Normalized qubit spectra obtained from Ramsey interferometry with the Kittel mode in (b)~the vacuum state and in (c)~a coherent state corresponding to a population of $\overline{n}_\mathrm{m}=0.53$ magnons obtained through a continuous coherent drive resonant with the Kittel mode. The shift per magnon $2\left|\chi_\mathrm{q-m}\right|/2\pi=2.5$~MHz is larger than the linewidth of each of the systems. The frequency of the Kittel mode with the qubit in the ground state, $\omega_\mathrm{m}^g/2\pi=7.7915$~GHz, is far detuned from the qubit frequency with the Kittel mode in the vacuum state, $\omega_\mathrm{q}^{0}/2\pi=7.9332$~GHz.
\label{fig:strong_dispersive_regime}}
\end{center}\end{figure}

Of particular interest is the strong dispersive regime, characterized by a shift per excitation larger than the linewidths of both systems~\cite{Gambetta2006,Schuster2007}. As illustrated in Fig.~\ref{fig:strong_dispersive_regime}(a), this regime enables one to resolve quanta of excitations in the bosonic mode to distinguish, for example, coherent, thermal, and squeezed states~\cite{Schuster2007,Kono2017}. Experimentally, the strong dispersive regime has been used in cQED to resolve single microwave photons in a mode of a microwave cavity~\cite{Schuster2007,Suri2015,Kono2017} and in cavity electromechanics to resolve single microwave-frequency phonons~\cite{Viennot2018,Arrangoiz-Arriola2019,Sletten2019}.

The strong dispersive regime of quantum magnonics was demonstrated in Ref.~\citenum{Lachance-Quirion2017} by performing qubit spectroscopy while driving the Kittel mode to create a steady-state coherent state with different average numbers of magnons $\overline{n}_\mathrm{m}$. Through a careful analysis of the spectra based on the theory of Ref.~\citenum{Gambetta2006}, these results establish that the Kittel mode has a thermal population below $0.01$~magnons, as expected from thermal equilibrium at $T=10$~mK. This shows that the Kittel mode is very close to the vacuum state when undriven; a significant advantage compared to MHz-frequency mechanical modes which need to be sideband cooled to reach the quantum ground state~\cite{Teufel2011a}. Furthermore, the probability distributions of the magnon number states $|n_\mathrm{m}\rangle$ follow closely the Poisson distributions expected for a bosonic mode~\cite{Rezende1969,Zagury1971}.

As an alternative to the continuous-wave measurements used in Ref.~\citenum{Lachance-Quirion2017}, the qubit absorption spectrum can be obtained from the Fourier transform of the Ramsey fringes measured in a time-domain experiment. Figure~\ref{fig:strong_dispersive_regime}(b) shows the resulting spectrum when the Kittel mode is undriven. Compared to the continuous-wave measurements, the broadening of the qubit by the spectroscopy tone is avoided and a qubit linewidth $\gamma_\mathrm{q}/2\pi=0.38$~MHz is obtained, corresponding to a coherence time $T_2^*=850$~ns, limited by the Purcell effect and dephasing from thermal photons in the microwave cavity modes. Figure~\ref{fig:strong_dispersive_regime}(c) shows the corresponding measurement in the presence of a coherent drive on resonance with the Kittel mode when the qubit is in the ground state at $\omega_\mathrm{m}^g/2\pi=7.7915$~GHz. A peak corresponding to a single magnon in the Kittel mode is clearly visible as the shift per magnon $2\left|\chi_\mathrm{q-m}\right|/2\pi=2.5$~MHz is larger than the linewidths of the Kittel mode and the qubit of $1.4$~MHz and $0.38$~MHz, respectively. As later discussed, the ability to count magnons, starting from the vacuum, is a key ingredient for future applications of quantum magnonics.

\subsection{Other nonlinearities}

\subsubsection{Self-Kerr effect}
The strong and coherent coupling between the Kittel mode and a superconducting qubit gives access to new types of nonlinearities in quantum magnonics beyond the dispersive qubit-magnon interaction previously discussed. Indeed, a self-Kerr nonlinearity at the level of single magnons is made possible by the presence of the qubit~\cite{Lachance-Quirion2017}. In the presence of such an interaction, the Hamiltonian of the magnetostatic mode becomes
\begin{align}
\mathcal{\hat H}_\mathrm{m}/\hbar=\left(\omega_\mathrm{m}+\frac{K_\mathrm{m}}{2}\right)\hat c^\dagger\hat c-\frac{K_\mathrm{m}}{2}\left(\hat c^\dagger\hat c\right)^2,
\end{align}
where $K_\mathrm{m}$ is the coefficient of the self-Kerr interaction~\cite{Nigg2012a,Bourassa2012,Kirchmair2013}. In Ref.~\citenum{Lachance-Quirion2017}, a self-Kerr coefficient $K_\mathrm{m}/2\pi=-0.2$~MHz is detected by the presence of a small nonlinearity in the average number of magnons $\overline{n}_\mathrm{m}$ when increasing the Kittel mode drive power. While this nonlinearity is smaller than the linewidth of the Kittel mode, it is many orders of magnitude larger than the intrinsic nonlinearity of the magnetostatic modes~\cite{Haigh2015,Wang2018} and could potentially be used to study nonlinear magnonics in the quantum regime. Furthermore, it is worth noting that it is possible to achieve a large qubit-magnon dispersive interaction with $\left|\chi_\mathrm{q-m}\right|\gg\gamma_\mathrm{m}$ while keeping the Kittel mode linear with $\left|K_\mathrm{m}\right|\ll\gamma_\mathrm{m}$ by a careful choice of the magnon frequency $\omega_\mathrm{m}$ relative to the qubit transition frequencies $\omega_\mathrm{q}$ and $\omega_\mathrm{q}+\alpha$~\cite{Juliusson2016,Lachance-Quirion2017}. 

\subsubsection{Cross-Kerr effect}
A cross-Kerr nonlinearity between the magnetostatic modes and the microwave cavity modes is also made possible by their mutual coupling to the qubit~\cite{Hu2011a,Holland2015}. In the presence of such a cross-Kerr interaction, the Hamiltonian of the interaction between a magnetostatic mode and a microwave cavity mode becomes
\begin{align}
\mathcal{\hat H}_\mathrm{m-c}/\hbar=g_\mathrm{m-c}\left(\hat a^\dagger\hat c+\hat a\hat c^\dagger\right)+K_\mathrm{m-c}\hat a^\dagger\hat a\hat c^\dagger\hat c,
\end{align}
where $K_\mathrm{m-c}$ is the cross-Kerr coefficient~\cite{Nigg2012a,Holland2015} which depends strongly on the magnon frequency relative to the qubit transition frequencies. The cross-Kerr interaction leads to a frequency shift of the microwave cavity mode (magnetostatic mode) for every magnon (microwave photon) excited in the magnetostatic mode (microwave cavity mode). Evidence of such a cross-Kerr interaction in quantum magnonics is seen in Ref.~\citenum{Lachance-Quirion2017} as a power-dependent offset of the qubit spectra, indicating a magnon-number-dependent shift of the readout cavity mode. As for the self-Kerr effect, the amplitude of the cross-Kerr effect depends strongly on the frequency of the Kittel relative to the transition frequencies of the qubit. In the single-excitation-resolved cross-Kerr interaction regime, corresponding to $\left|K_\mathrm{m-c}\right|\gg\kappa_\mathrm{c},\gamma_\mathrm{m}$, a number state in one of the modes can be stabilized through dissipation in the other mode~\cite{Holland2015}.

\section{Cavity optomagnonics}
\label{sec:optomagnonics}

Similar to the interaction between magnetostatic modes and superconducting qubits, the interaction between light and spin waves in ferromagnetic crystals is indirect. Indeed, the optomagnonic interaction consists of electric dipole interactions between the electric component of the optical field and the electrons of the ferromagnetic material, mediated by the spin-orbit interaction~\cite{Elliott1963,Shen1966,Moriya1968,LeGall1971}. The magnetic dipole interaction, in comparison, is negligible~\cite{Bass1960,Shen1967,Fleury1968}. Phenomenologically, magneto-optical effects are contained in the expression of the dielectric tensor $\tilde\epsilon(\mathbf{M})$ which depends on the magnetization $\mathbf{M}$~\cite{Moriya1968,Pershan1967,Wettling1976} and encapsulates the Faraday and Cotton-Mouton effects. For decades, light has been used to probe magnon physics. As illustrated in Fig.~\ref{fig:optomagnonics}~(a), now with the advent of quantum magnonics, the coherent control of spin waves with light holds the promise for an efficient microwave-to-optics transduction, enabling quantum-limited microwave amplification and optical interfacing of superconducting qubits. After the first implementation of such a magnon-based transducer~\cite{Hisatomi2016}at the frontiers between optomagnonics and cavity electromagnonics~[Fig.~\ref{fig:optomagnonics}(b)], the interaction is now explored in an optical cavity, revealing a new physical playground, cavity optomagnonics [Fig.~\ref{fig:optomagnonics}(c)].

\subsection{Optomagnonics in solids}

\subsubsection{Light as a probe}
From pioneering experiments revealing one and two-magnon scattering in the antiferromagnetic material FeF$_2$~\cite{Fleury1966}, coherent spin waves in a YIG rectangular bar~\cite{Hu1971} and thermally-excited magnons in metallic thin films of Ni and Fe~\cite{Sandercock1978}, light scattering has been frequently used to probe spin-wave physics. Micro-focused Brillouin light scattering~\cite{Sebastian2015}, with a diffraction-limited spatial resolution, is now a mature imaging tool for magnon spintronics~\cite{Demidov2008,Vogt2014} and fundamental magnon studies~\cite{Mathieu1998,Schultheiss2008,Sandweg2011}, notably for the observation of Bose-Einstein condensation of magnons at room temperature in YIG thin films~\cite{Demokritov2006,Serga2014}.

\begin{figure}\begin{center}
\includegraphics[scale=1]{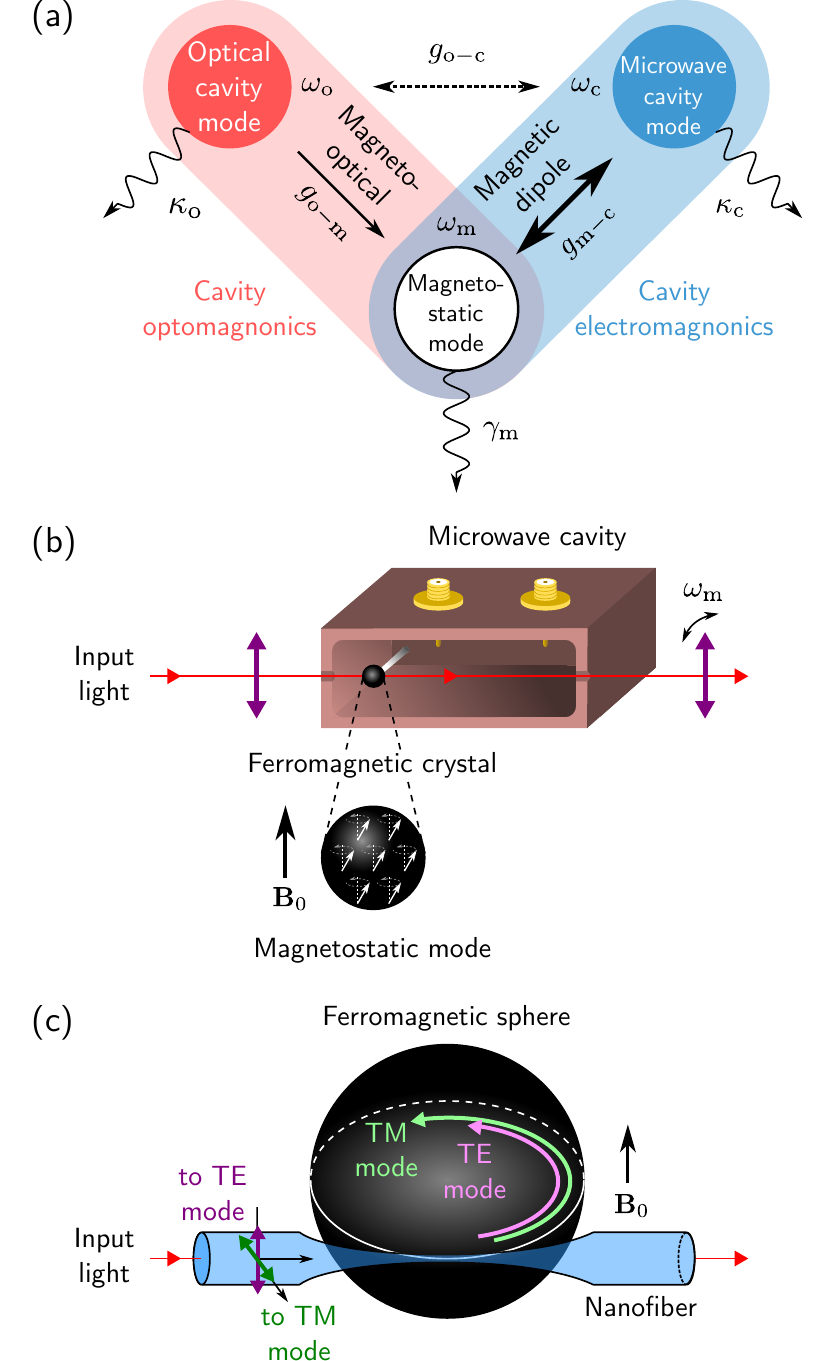}
\caption{\textbf{Cavity optomagnonics.}
(a)~Schematic diagram of the optomagnonic coupling $g_\mathrm{o-m}$ between a magnetostatic mode and an optical cavity mode of frequencies $\omega_\mathrm{m}$ and $\omega_\mathrm{o}$, respectively. The magnetostatic mode can be furthermore coupled to a microwave cavity mode of frequency $\omega_\mathrm{c}$ through a magnetic dipole interaction of coupling strength $g_\mathrm{m-c}$. The combination of cavity optomagnonics and cavity electromagnonics therefore enables an effective coupling $g_\mathrm{o-c}$ between optical and microwave cavity modes. The linewidths of the optical cavity mode, the magnetostatic mode and the microwave cavity modes are respectively $\kappa_\mathrm{o}$, $\gamma_\mathrm{m}$, and $\kappa_\mathrm{c}$, respectively.
(b)~Schematic representation of a device realizing the architecture of cavity electromagnonics probed with light. Through the Faraday effect, the polarization of optical photons passing through the microwave cavity and the ferromagnetic crystal will be modulated at the frequency of the magnetostatic mode $\omega_\mathrm{m}$ when magnons are excited.
(c)~Schematic representation of light evanescently coupled to whispering gallery modes in a ferromagnetic sphere through a nanofiber along the plane transverse to the external magnetic field $\mathbf{B}_0$, such that whispering gallery and magnetostatic modes share the same equatorial plane. Panel~(c) is adapted from Ref.~\citenum{Osada2016}.
\label{fig:optomagnonics}}
\end{center}\end{figure}

\subsubsection{Light as a control interface}
Another attractive aspect of optomagnonics lies in the ability to control magnons with light. For example, ultrafast optics experiments have unveiled the time-resolved dynamics of the magnetization in nanostructures by exciting spin waves with femtosecond pulsed lasers~\cite{Acremann2000,VanKampen2002,Kirilyuk2010,Satoh2012}. Later, the bidirectional and coherent conversion between optical and microwave photons using spin waves was demonstrated with the Kittel mode of a $750~\mu$m-diameter YIG sphere~\cite{Hisatomi2016}. As schematically illustrated in Fig.~\ref{fig:optomagnonics}(b), the sphere is embedded in a three-dimensional microwave cavity and is illuminated by a $1550$~nm continuous-wave laser. The Kittel mode is hybridized with the fundamental mode of the microwave cavity through a magnetic dipole coupling. Induced by the spin waves excited by the microwave photons, the Faraday effect results in a modulation of the polarization of the transmitted beam at the frequency of the hybridized mode, generating two optical sidebands around the laser frequency which can be detected by a heterodyne measurement with a high-speed photodiode.

The creation of microwave photons by light is, on the other hand, established by using two co-propagating lasers with a stabilized relative phase and frequencies different by the magnon frequency. The created magnons relax in the microwave cavity, resulting in a microwave signal measured with a vector network analyzer. The phase coherence is preserved in the conversion process, resulting in a photon conversion efficiency of $\sim10^{-10}$ for $20$~mW of optical pump power. Bidirectional conversion between microwave and optical photons is appealing for quantum-limited microwave amplifiers~\cite{Barzanjeh2015} and quantum telecommunications~\cite{Kimble2008,Wehner2018}. This inaugural work calls for more developments to increase the optomagnonic interaction, especially towards experiments in an optical cavity. 

\subsection{Cavity optomagnonics in solids}

\subsubsection{Optical whispering gallery modes}
A spherical dielectric ferromagnetic crystal itself hosts optical whispering gallery modes (WGMs)~\cite{Haigh2015}. Optical WGMs in monolithic resonators offering high optical quality factors and relatively small mode volumes~\cite{Braginsky1989,Armani2003} are particularly appealing for cavity quantum electrodynamics~\cite{Aoki2006}, optomechanics~\cite{Schliesser2008,Matsko2009}, and frequency combs~\cite{DelHaye2007}. The low absorption of YIG in the telecommunication band~\cite{Wood1967} could lead to optical quality factors of $\sim10^6$, but has been limited to an order of magnitude less by losses due to surface roughness, resulting in an optical cavity linewidth of $\kappa_\mathrm{o}/2\pi\sim2$~GHz. In a large sphere, the optical WGMs are confined close to the equator and can be discriminated into two families; quasi-TM and quasi-TE modes, corresponding respectively to a polarization in-plane and out-of-plane with respect to the equator. The input and output light modes are evanescently coupled to the optical cavity along the plane transverse to the external magnetic field $\mathbf{B}_0$, such that whispering gallery and magnetostatic modes share the same equatorial plane [Fig.~\ref{fig:optomagnonics}(c)].

Similar to Brillouin light scattering in thin films, an incoming photon is scattered by a magnon into an orthogonally polarized photon. The scattered photon sideband is dependent on whether the process creates or annihilates a magnon. The Hamiltonian of the linearized optomagnonic interaction can be written as
\begin{align}
\mathcal{\hat H}_\mathrm{o-m}&=\hbar g_\mathrm{o-m}\left(\hat b^\dagger\hat c+\hat b\hat c^\dagger\right),\\
&=\frac{1}{2}\int\mathrm{d}\mathbf{r}\ \mathrm{d}t\ \mathbf{E}_\mathrm{out}^*(t)\tilde\epsilon(\mathbf{M})\mathbf{E}_\mathrm{in}(t),
\end{align}
where $g_\mathrm{o-m}$ is the optomagnonic coupling strength, $\hat b^\dagger$ and $\hat b$ respectively creates and annihilates an optical photon from the WGM, and $\mathbf{E}_\mathrm{in}(t)$ and $\mathbf{E}_\mathrm{out}(t)$ are respectively the electric fields of the input and scattered photons of the WGM~\cite{Hisatomi2016,Osada2016,Zhang2016,Haigh2016,Kusminskiy2016}. The contribution of the spin-wave mode is enclosed in the permittivity $\tilde\epsilon(\mathbf{M})$ through the expression of its magnetization $\mathbf{M}$. The scattered light leaks out of the cavity and can be measured by a heterodyne measurement with a high-speed photodiode.

\subsubsection{Triple resonance condition}
By energy conservation, the optomagnonic interaction should be maximized at the triple resonance condition, where the frequency of the considered magnetostatic mode matches the difference between the frequencies of the input and scattered optical photons. Two optical modes of the same polarization family separated by one azimuthal index are split by the free spectral range $\mathrm{FSR}=c/\pi dn_\mathrm{YIG}$, where $c$ is the speed of light, $d$ is the diameter of the sphere, and $n_\mathrm{YIG}=2.19$ is the refractive index of YIG. For a millimeter-sized sphere, the resulting $\mathrm{FSR}$ is about $50$~GHz. Due to birefringences in the sphere, the frequency of a TM mode is systematically shifted by $B\sim\mathrm{FSR}\sqrt{n_\mathrm{YIG}^2-1}/n_\mathrm{YIG}\approx0.89\times\mathrm{FSR}$ compared to a TE mode with the same indices. Therefore, for an external magnetic field $B_0\approx0.3$~T, leading to magnetostatic mode frequencies in the GHz range, the triple resonance condition is within reach for scattering between a TM mode and a TE mode separated by one azimuthal index, as their frequency difference is $\mathrm{FSR}-B\approx0.11\times\mathrm{FSR}\sim5$~GHz. 

\subsubsection{Experiments on the Kittel mode}
In an early experiment, infrared light was evanescently coupled with a tapered fiber to the optical modes of a $750~\mu$m-diameter YIG sphere~\cite{Osada2016}. This first demonstration of Brillouin light scattering of WGM photons with magnons in the Kittel mode revealed non-reciprocal behavior: the two possible input orbit directions of light in the sample, with respect to the external magnetic field, resulted in scattering levels which differed by one order of magnitude. 

Demonstration of the triple resonance has been realized on spheres with diameters of $500~\mu$m and $300~\mu$m, respectively coupled by a birefringent rutile prism~\cite{Haigh2016} and a SiN optical waveguide~\cite{Zhang2016}. These experiments are deep into the weak coupling regime as the intrinsic optomagnonic coupling strength $g_\mathrm{o-m}/2\pi\sim5$~Hz is much smaller than the linewidths of both the magnetostatic and optical cavity modes, $\gamma_\mathrm{m}/2\pi\sim1$~MHz and $\kappa_\mathrm{o}/2\pi\sim1$~GHz~\cite{Haigh2016,Zhang2016}, respectively. Following these early demonstrations, a theoretical framework has been established for cavity optomagnonics~\cite{Kusminskiy2016,Liu2016b,Pantazopoulos2017}, followed by thorough work more specifically on spherical cavity optomagnonics, extending the studies to spin-wave modes beyond the Kittel mode~\cite{Sharma2017}. 

\subsubsection{Higher-order magnetostatic modes and angular momentum conservation}
Higher-order magnetostatic modes, presenting a variety of spin textures, allow a deeper understanding of the optomagnonic interaction. In particular, the exchange of orbital angular momentum between magnons and optical photons has been experimentally demonstrated with a $1$~mm-diameter YIG sphere~\cite{Osada2018,Haigh2018}. Associated with birefringences in the system favoring a particular sideband, the selection rules along the azimuthal axis explain the non-reciprocal behavior of the Brillouin light scattering. Dependent on the azimuthal dependence of the considered magnetostatic mode and on the input orbit direction of light in the sample~\cite{Osada2017}, this phenomenon constitutes an original implementation of chiral photonics~\cite{Lodahl2017}. 

\subsection{Future directions}    
  
\subsubsection{Increasing the optomagnonic coupling}
The optomagnonic coupling, faint for the Kittel mode because of the low spatial overlap with the optical WGMs, will be increased for higher-order modes whose spatial distribution is further localized towards the resonator boundaries where the optical modes are located~\cite{Osada2018,Haigh2018}. The challenge lies in the ability to efficiently excite these higher-order modes, respecting energy and orbital angular momentum conservation laws and eventually being able to identify them properly~\cite{Gloppe2018}.

New magnetic materials could be investigated, such as bismuth-doped YIG~\cite{Lacklison1973} and, at low temperatures, CrBr$_3$~\cite{Dillon1963} or fully concentrated rare-earth ion crystals~\cite{Everts2018}. The ideal material would require a small linewidth $\gamma_\mathrm{m}$ for the magnetostatic modes, strong magneto-optical effects, low optical absorption, and the technological ability to prepare surfaces which allow for high optical quality factors. The magnetostatic mode volume could also be tuned by changing the shape of the optical resonator. A disk would have reduced spin-wave mode volumes compared to a sphere of the same diameter. Additionally, as recently proposed in Ref.~\citenum{Graf2018}, this geometry would allow the exploration of a new kind of optomagnonic coupling by working with a vortex mode in an optical resonator, which could also couple to microwave-frequency cavity modes through a magnetic dipole interaction~\cite{Martinez-Perez2018}.

More generally, the size of the sample can be reduced to increase the optomagnonic coupling~\cite{Kusminskiy2016}. The challenge here lies in preparing high optical quality magnetic structures below $250~\mu$m. Recent progress in YIG microfabrication~\cite{Seo2017,Heyroth2018} could grant access to regimes where light-matter interactions are governed by optical Mie resonances~\cite{Almpanis2018}, potentially opening up a new paradigm of cavity nano-optomagnonics. For quantum transducers based on magnonics, the reduction of the size of the ferromagnetic crystal should be explored while keeping the magnetic-dipole interaction with the microwave cavity modes in the strong coupling regime using, for example, microwave cavities with smaller mode volumes~\cite{McKenzie-Sell2019}.

\subsubsection{Perspectives with stronger coupling}
Cavity cooling and amplification of magnons by light has been studied theoretically~\cite{Sharma2018} and will be within experimental reach if the optomagnonic coupling can be increased by about two orders of magnitude. Analogous to the DLCZ protocol~\cite{Duan2001}, a heralding protocol consisting of the creation of magnon Fock states through entanglement with optical photons has been recently theoretically investigated ~\cite{Bittencourt2018}. In the strong coupling regime, this would constitute an important step towards quantum information protocols in optomagnonic devices. The field of cavity optomagnonics in the solid state is just emerging and is expected to experience a blooming of new implementations in the near future. 

\section{Cavity magnomechanics}
\label{sec:magnomechanics}

\begin{figure}\begin{center}
\includegraphics[scale=1]{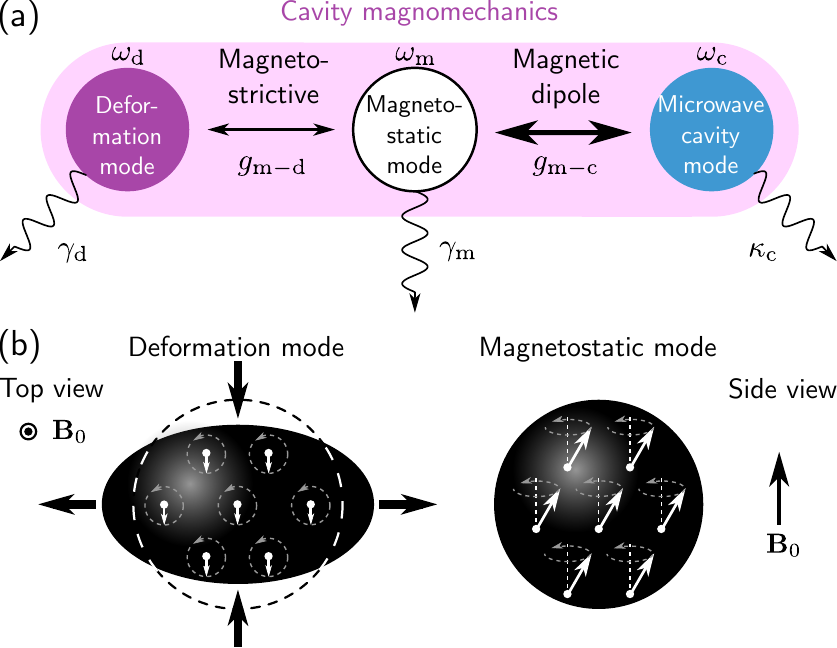}
\caption{\textbf{Cavity magnomechanics.}
(a)~Schematic diagram of the coupling between a deformation mode, a magnetostatic mode, and a microwave cavity mode of frequencies $\omega_\mathrm{d}$, $\omega_\mathrm{m}$, and $\omega_\mathrm{c}$, respectively. The magnetostrictive force leads to a coupling between deformation and magnetostatic modes of strength $g_\mathrm{m-d}$, which can be parametrically enhanced beyond the linewidth $\gamma_\mathrm{d}$ of the mechanical mode by driving the magnetostatic mode. A magnetic dipole interaction of coupling strength $g_\mathrm{m-c}$ further couples the magnetostatic and microwave cavity modes. (b)~Schematic representation of the coupling between a deformation mode (left) and a magnetostatic mode (right) enabled through magnetostrictive forces.
\label{fig:magnomechanics}}
\end{center}\end{figure}

In analogy to cavity optomechanics~\cite{Aspelmeyer2014} and cavity electromechanics~\cite{Regal2011}, the use of mechanical degrees of freedom in ferromagnetic crystals is a natural new avenue for hybrid quantum systems based on magnonics~\cite{Zhang2015,Li2018}. Notably, deformation modes in a ferromagnetic crystal could be used as mechanical modes intrinsic to the sample that couples to magnetostatic modes through magnetostrictive forces~\cite{Spencer1958,Schlomann1960}. As depicted in Fig.~\ref{fig:magnomechanics}(a), combined with the magnetic dipole interaction between magnetostatic and microwave cavity modes, this leads to the recently demonstrated platform of cavity magnomechanics~\cite{Zhang2015}.

\subsubsection{Coupling to deformation modes}

A mechanical mode of frequency $\omega_\mathrm{d}$ can be described as a harmonic oscillator by the Hamiltonian
\begin{align}
\mathcal{\hat H}_\mathrm{d}/\hbar=\omega_\mathrm{d}\hat d^\dagger\hat d,
\end{align}
where $\hat d^\dagger$ and $\hat d$ respectively creates and annihilates a phonon in the mechanical mode. When this mode corresponds to a deformation mode hosted in a ferromagnetic material, magnetostrictive forces lead to a radiation pressure-like interaction between a magnetostatic mode and the mechanical mode described by the Hamiltonian
\begin{align}
\mathcal{\hat H}_\mathrm{m-d}/\hbar=g_\mathrm{m-d}\hat c^\dagger\hat c\left(\hat d+\hat d^\dagger\right),
\end{align}
where $g_\mathrm{m-d}$ is the magnomechanical coupling strength~\cite{Zhang2015}. As opposed to the beam splitter-like interaction previously discussed, the radiation pressure-like interaction between the deformation and magnetostatic modes enables the access, as in cavity optomechanics~\cite{Aspelmeyer2014}, to phenomena such as sideband cooling of the mechanical mode and parametric enhancement of the coupling strength.

\subsubsection{Experiment}
The platform of cavity magnomechanics has been first demonstrated at room temperature with a millimeter-sized YIG sphere inside a three-dimensional microwave cavity~\cite{Zhang2015}. For spheres with a diameter of approximately $250~\mu$m, the frequency $\omega_\mathrm{d}$ of the low-order mechanical modes reaches about $10$~MHz. Mechanical modes at larger frequencies could be possible by using micrometer-sized ferromagnetic mechanical oscillators~\cite{Seo2017,Heyroth2018}. In the experiment of Ref.~\citenum{Zhang2015}, in order to greatly reduce the clamping losses of the deformation modes, the YIG sphere is glued to an optical fiber, enabling a mechanical linewidth of $\gamma_\mathrm{d}/2\pi=150$~Hz.

In this system, the magnomechanical coupling strength $g_\mathrm{m-d}/2\pi\sim10$~mHz is much smaller than every linewidth in the hybrid system. However, as the magnetostrictive interaction is of the radiation pressure type, the coupling strength can be parametrically enhanced to approximately $30$~kHz by strongly pumping the magnetostatic mode~\cite{Zhang2015}. This enhancement leads to a coupling strength larger than the linewidth $\gamma_\mathrm{d}$ of the deformation mode, but still smaller than the $\sim$MHz linewidth of magnetostatic modes in YIG spheres. Despite this limitation, a large diversity of phenomena has been observed in this hybrid system~\cite{Zhang2015}. Notably, the so-called triple resonance condition, in which the frequency of the mechanical mode matches the magnetic dipole coupling strength between the resonant magnetostatic and microwave cavity modes, is demonstrated.

\subsubsection{Future directions}
Recently, it has been proposed that tripartite entanglement between the deformation, magnetostatic and microwave cavity modes could be realized in cavity magnomechanics to study, for example, macroscopic quantum phenomena~\cite{Li2018,Li2018b}. The key ingredient of the proposal is to strongly pump the magnetostatic mode in order to enhance the magnostrictive interaction and to cool down the mechanical mode to the quantum ground state using sideband cooling~\cite{Aspelmeyer2014}. It is shown that the multipartite entanglement survives for temperatures slightly above $\sim100$~mK. This proposal could be challenging to achieve experimentally as it requires one to pump the magnetostatic mode with milliwatts of power, a feat potentially difficult to achieve in a dilution refrigerator. Despite the possible difficulties, this proposal shows the potential of cavity magnomechanics to bring quantum phenomena to magnonics.

\section{Perspectives and outlook for quantum magnonics}
\label{sec:outlook}

Hybrid systems based on magnonics have progressed at a rapid pace since the first experiments were conducted~\cite{Huebl2013,Tabuchi2014,Tabuchi2015,Lachance-Quirion2017,Hisatomi2016,Osada2016,Osada2018,Haigh2016,Zhang2015}. This progress has been partially made possible by the adaption of tools developed in cavity and cQED to the field of quantum magnonics. This path has enabled steady progress, while the features specific to quantum magnonics, which can be considered as quantum optics in magnetically-ordered solid-state systems, still provide insight into newly discovered physical phenomena and novel applications.

\begin{figure}\begin{center}
\includegraphics[scale=1]{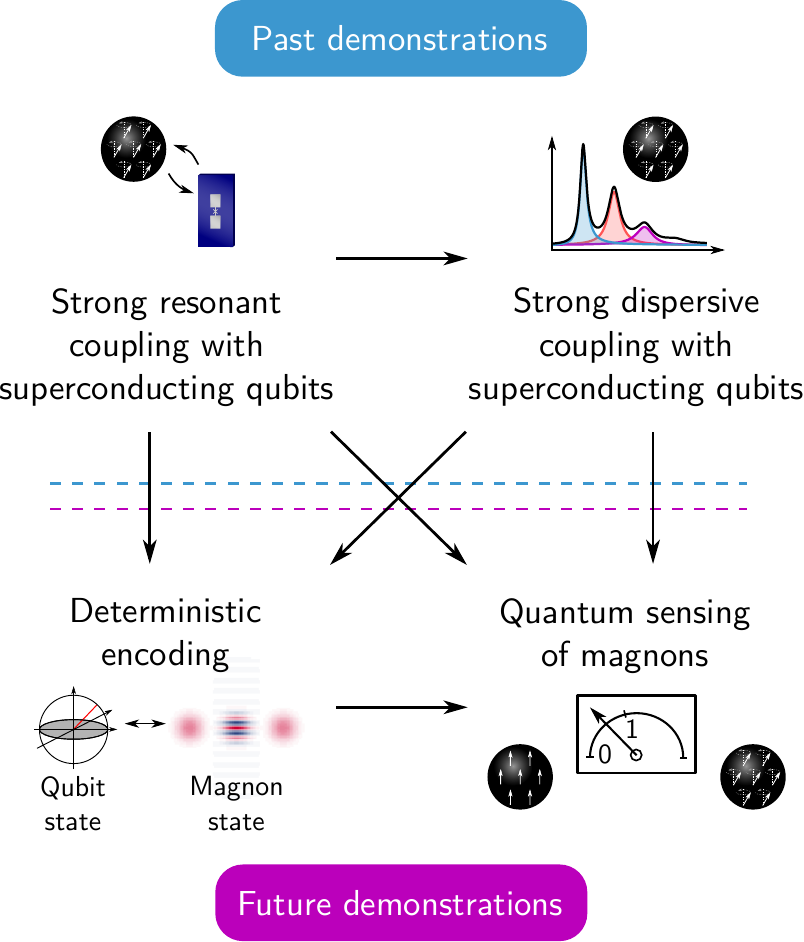}
\caption{\textbf{Perspectives and outlook for quantum magnonics.}
Schematic diagram of past and future demonstrations of quantum magnonics. The strong resonant coupling of magnetostatic modes to superconducting qubits has enabled one to reach the strong dispersive coupling. Both regimes should be useful for potential future demonstrations, such as encoding arbitrary qubit states into nonclassical magnon states and quantum sensing of magnons.
\label{fig:perspectives}}
\end{center}\end{figure}

The important steps demonstrated in quantum magnonics, as well as potential future demonstrations, are shown schematically in Fig.~\ref{fig:perspectives}. Two main applications are envisioned here. The first one aims to develop a quantum transducer based on magnonics to interface microwave-only superconducting circuits with optical photons~\cite{Kimble2008,Kurizki2015,Hisatomi2016,Zhong2019}. One key ingredient for such a technology to be demonstrated in quantum magnonics is to faithfully transfer quantum information from a superconducting qubit to a magnetostatic mode. The second main application aims to use quantum magnonics to provide sensitive detection of magnons through quantum-enhanced sensing protocols~\cite{Degen2017}. As later discussed, this could be useful for a variety of fields, from ultra-sensitive detection of magnetic excitations for magnon spintronics to dark-matter searches for axions.

\subsection{Encoding quantum information in magnetostatic modes}
As a magnetostatic mode can be described as a harmonic oscillator, arbitrary nonclassical states are not naturally created in this system. Indeed, for example, a coherent drive creates coherent states of magnons~\cite{Rezende1969,Lachance-Quirion2017}. A nonlinearity in the system is therefore necessary in order to create nonclassical states. Furthermore, while creating any nonclassical state in a magnetostatic mode is a great feat in itself, schemes enabling one to encode arbitrary quantum states of a qubit into the quantum state of a magnetostatic mode are of potentially even greater interest. We identify two such schemes based on the strong resonant and dispersive couplings between a superconducting qubit and a magnetostatic mode previously demonstrated in quantum magnonics~\cite{Tabuchi2015,Lachance-Quirion2017}.

\subsubsection{Schemes based on a resonant coupling}
The most natural scheme to prepare quantum states in a magnetostatic mode is by using the strong and coherent interaction with a superconducting qubit when both systems are resonant~\cite{Tabuchi2015}. Indeed, on resonance, quanta of excitations are exchanged between the two systems at a rate proportional to the coupling strength $g_\mathrm{q-m}$. Therefore, a single excitation can be transferred to the magnetostatic mode after preparing the qubit in the excited state $|e\rangle$ using a classical drive, an operation possible due to the nonlinearity of the qubit. This scheme has been successfully implemented to create quantum states of the motional degree of freedom of a trapped ion~\cite{Meekhof1996}, photons in a microwave resonator~\cite{Hofheinz2008,Hofheinz2009} and, more recently, phonons in a mechanical oscillator~\cite{Satzinger2018,Chu2018}.

With an always-on resonant interaction, the exchange of excitations is continuous and a single excitation is swapped back and forth between the two systems. It is therefore necessary to have either a dynamically tunable coupling or detuning. With such a dynamical control, after the excitation is transferred from the qubit to the bosonic mode, the coupling can be turned off to halt the exchange, or, alternatively, the detuning can be increased to suppress the amplitude of the exchange. The quantum states naturally created with this resonant scheme are number states~\cite{Hofheinz2008,Satzinger2018,Chu2018} as well as arbitrary quantum states~\cite{Law1996,Hofheinz2009}.

A dynamically tunable coupling has been demonstrated in quantum magnonics~\cite{Tabuchi2015}, and a dynamically tunable detuning could be implemented, for example, by using a flux-tunable superconducting qubit~\cite{Hofheinz2008,Hofheinz2009,Satzinger2018,Chu2018}. Therefore, arbitrary states of the qubit can be encoded in a magnetostatic mode provided that the coupling strength $g_\mathrm{q-m}$ is much larger than the rate at which the quantum information is lost in the qubit ($\gamma_\mathrm{q}=2/T_2^*$) and in the magnetostatic mode ($\gamma_\mathrm{m}$). While the strong coupling regime of quantum magnonics has been demonstrated~\cite{Tabuchi2015}, improvements are required to further increase the ratio between the coupling strength and the decoherence and dissipation rates.

Two possible paths for increasing the coupling strength $g_\mathrm{q-m}$ relative to the linewidths of the different constituents of the hybrid system are discussed here. First, in order to decrease the qubit linewidth to a few tens of kHz, the losses of the microwave cavity modes need to be reduced. However, as an external magnetic field needs to be applied to the ferromagnetic crystal inside the microwave cavity, this cannot be readily achieved by using a standard superconducting microwave cavity due to the Meissner effect. Secondly, as there is no clear path on how to reduce the linewidth of magnetostatic modes significantly below $1$~MHz in the quantum regime~\cite{Tabuchi2014}, the coupling strength between the magnetostatic modes and the qubit needs to be increased from the $\sim10$~MHz previously demonstrated~\cite{Tabuchi2015,Tabuchi2016,Lachance-Quirion2017}. This can be achieved, for example, by increasing the strength of the magnetic dipole interaction between the magnetostatic modes and the microwave cavity modes. Such improvements are within experimental reach as magnetic dipole coupling strengths up to $2$~GHz have been demonstrated in cavity electromagnonics~\cite{Goryachev2014a}.

\subsubsection{Schemes based on a dispersive coupling}
As an alternative to the resonant coupling, a dispersive interaction can be used to transfer the quantum information encoded in superconducting qubits into bosonic modes such as magnetostatic modes~\cite{Haroche2006,Vlastakis2013a}. An advantage of this approach is that a dynamical control of the coupling or the detuning is not necessary. However, the time required to perform the encoding is proportional to $1/\left|\chi_\mathrm{q-m}\right|$, as opposed to $1/g_\mathrm{q-m}$ for the resonant scheme, such that any scheme based on a dispersive coupling is inherently slower than those based on a resonant coupling. Another distinction is that the quantum states prepared using a dispersive interaction are superpositions of coherent states, often called cat states.

The protocol used in Ref.~\citenum{Vlastakis2013a} to encode the quantum state of a superconducting qubit into a cat state in a microwave cavity mode will be described as an illustrative example. This scheme uses the strong dispersive interaction between the qubit and the bosonic mode enabling conditional operations on one system depending on the state of the other system. More specifically, a coherent state in the bosonic mode acquires an additional phase shift if the qubit is in the excited state $|e\rangle$, therefore corresponding to a conditional rotation in phase space. Starting with the qubit in a coherent superposition of the ground and excited states, this operation directly leads, after a free evolution time $\tau=\pi/2\left|\chi_\mathrm{q-m}\right|$, to a cat state in the bosonic mode entangled with the qubit. Additional steps enable one to disentangle the bosonic mode and the qubit. In the end, deterministic encoding of arbitrary quantum states of the qubit into cat states of the bosonic mode can be achieved with this protocol~\cite{Vlastakis2013a}.

While the strong dispersive regime has been previously demonstrated in quantum magnonics~\cite{Lachance-Quirion2017}, the ratio between the amplitude of the dispersive shift $\left|\chi_\mathrm{q-m}\right|$ and the linewidths of the Kittel mode and the qubit, $\gamma_\mathrm{m}$ and $\gamma_\mathrm{q}$ respectively, needs to be significantly improved in order to implement the encoding protocol presented here. This can be quantified by the dispersive cooperativity $C_\chi\equiv4\left|\chi_\mathrm{q-m}\right|^2/\left(\gamma_\mathrm{m}\gamma_\mathrm{q}\right)$. In quantum magnonics, the best demonstrated dispersive cooperativity was $C_\chi\sim12$~\cite{Lachance-Quirion2017}. In comparison, in cQED, $C_\chi$ reaches $\sim8\times10^4$~\cite{Vlastakis2013a}. While the amplitude of the dispersive interaction is similar, the dispersive cooperativity is much larger in the latter thanks to the very small linewidth of three-dimensional superconducting microwave cavity modes ($\kappa_\mathrm{c}/2\pi\sim10$~kHz) compared to the linewidth of magnetostatic modes ($\gamma_\mathrm{m}/2\pi\sim1$~MHz). Furthermore, in quantum magnonics, the qubit linewidth $\gamma_\mathrm{q}/2\pi\approx0.4$~MHz is limited, at the moment, by a reduced lifetime due to the Purcell effect and increased dephasing due to thermal populations in the lossy microwave cavity modes mediating the coupling. For example, in order to reach the desired dispersive cooperativity $C_\chi=10^4$, the amplitude of the dispersive coupling needs to reach about $10$~MHz and the qubit linewidth needs to be reduced to about $40$~kHz. As both are realistic figures, the parameters necessary to faithfully encode the quantum state of a superconducting qubit into a nonclassical magnon state in a magnetostatic mode for quantum transducers should be within experimental reach in the near future.

\subsection{Quantum sensing of magnons}
The development of hybrid quantum systems based on magnonics opens up opportunities for using tools developed in cryogenic microwave experiments for quantum-enhanced detection of magnons. Sensing of magnons in hybrid quantum systems has been demonstrated in cavity electromagnonics at millikelvin temperatures. For example, ferromagnetic resonance has been measured while probing the system with, on average, much less than a single magnon in the ferromagnetic crystal~\cite{Tabuchi2014,Morris2017}. Magnetostatic modes can therefore be probed with minimal disturbance of the ferromagnetic order. For example, a new relaxation mechanism of magnons, which can only be observed in the quantum regime where the average number of excited magnons is much smaller than one, has been identified~\cite{Tabuchi2014}.

The detection of magnons at cryogenic temperatures has recently found applications in dark matter searches for galactic axions through a coupling to electron spins in a ferromagnetic material~\cite{Barbieri1989,Barbieri2017,Crescini2018a,Flower2018}. Indeed, according to a particular model of axions, the movement of Earth via the motion of the Solar System through the galactic halo of axions leads to an effective microwave-frequency magnetic field which excites magnons in the Kittel mode of a ferromagnetic crystal if the axion mass is such that it is resonant with this mode~\cite{Barbieri2017}. Preliminary experiments use strongly coupled and resonant magnetostatic and microwave cavity modes in the architecture of cavity electromagnonics to detect magnons possibly generated by the axion wind~\cite{Barbieri2017,Crescini2018a,Flower2018}. These experiments could be improved by using quantum-limited amplifiers such as narrowband Josephson parametric amplifiers~\cite{Yamamoto2008} or broadband traveling-wave parametric amplifiers~\cite{Macklin2015}, or, alternatively, by using single microwave photon detectors based on superconducting circuits~\cite{Inomata2016,Kono2018,Besse2018}.

The coupling of magnetostatic modes to superconducting qubits in quantum magnonics also enables quantum-enhanced sensing of magnons. Indeed, magnon number states $|n_\mathrm{m}\rangle$ have been resolved by using the qubit as a quantum probe~\cite{Lachance-Quirion2017}. This demonstration opens up the possibility to detect a population of magnons in the magnetostatic modes using quantum sensing protocols~\cite{Degen2017}. For example, the shift of the qubit frequency caused by a magnon population $\overline{n}_\mathrm{m}$ in a magnetostatic mode dispersively coupled to a qubit can be measured through Ramsey interferometry. Furthermore, in analogy to experiments in cQED~\cite{Narla2016}, the strong dispersive regime of quantum magnonics makes it possible to use the device as a single-magnon detector. The many avenues for quantum sensing of magnons in quantum magnonics, made possible by the coupling of the magnetostatic modes to the qubit, makes this architecture promising for achieving the equivalent of a single-photon detector for magnonics, a technological step potentially useful, for example, for magnon spintronics~\cite{Karenowska2015}.

\section{Conclusions}
\label{sec:conclusion}

To conclude, the various hybrid quantum systems based on magnonics outlined in this review article have allowed proof-of-principle experiments that show great potential for a variety of quantum technologies such as quantum-enhanced sensing of magnons and microwave-to-optical quantum transduction for superconducting circuits. More specifically, the development of cavity electromagnonics has prompted a large variety of experiments for the study of the coupling between collective spin excitations in different ferromagnetic materials and modes of different types of microwave cavities~\cite{Huebl2013,Tabuchi2014,Goryachev2014a,Zhang2014c}. Furthermore, we note that ferromagnetic crystals in microwave cavities enables one to study topological many-body states in arrays of microwave cavities with a nonreciprocity induced by ferromagnetic crystals~\cite{Anderson2016,Owens2018}.

Starting from cavity electromagnonics, the platform of quantum magnonics adds nonlinear elements in the form of superconducting qubits in order to engineer an effective strong and coherent interaction between qubits and magnetostatic modes~\cite{Tabuchi2015,Tabuchi2016}. While strong and coherent couplings have been demonstrated in both the resonant~\cite{Tabuchi2015} and dispersive~\cite{Lachance-Quirion2017} regimes, improvements in the microwave cavity should increase the coupling strengths and reduce qubit losses. Through these improvements, deterministic encoding of the state of a superconducting qubit into a nonclassical magnon state should be within experimental reach and would constitute a great step towards building a quantum transducer based on this architecture. Even with the currently demonstrated parameters, it is expected that quantum-enhanced sensing of magnons by using the qubit as a probe is possible, as best exemplified by the observation of magnon number states~\cite{Lachance-Quirion2017}. Bringing to the field of magnonics the equivalent of the single-photon detector for optics would be a great demonstration of the possibilities of quantum sensing in quantum magnonics.

In combination with cavity electromagnonics, cavity optomagnonics serves as a key ingredient for the bidirectional conversion between microwave and optical photons for the development of quantum transducers~\cite{Hisatomi2016}. Increasing the optomagnonic coupling with, for example, smaller ferromagnetic samples or higher-order magnetostatic modes, could potentially enable the efficient coherent control of magnons with light~\cite{Hisatomi2016,Kusminskiy2016}. Finally, the platform of cavity magnomechanics promises to offer the possibility to observe quantum effects in magnonics with linear systems only~\cite{Zhang2015,Li2018,Li2018b}. Overall, the diversity of platforms based on hybrid quantum systems with magnetostatic modes provides many opportunities for novel quantum technologies with applications in both fundamental and applied physics.

\acknowledgments
The authors would like to thank Samuel Piotr Wolski for performing some of the measurements presented here, and Nicolas Gheeraert and Jacob Koenig for carefully reading the manuscript. This work is partly supported by the Project for Developing Innovation System of MEXT, JSPS KAKENHI (26220601, 18F18015), JST ERATO (JPMJER1601), and FRQNT Postdoctoral Fellowships. D.L.-Q. is an International Research Fellow of JSPS.

\bibliography{library}
\bibliographystyle{apsrev4-1}

\end{document}